\documentclass[modern]{aastex631}

\usepackage{textcomp}

\accepted{May 4, 2021}
\submitjournal{Planetary Science Journal}

\shorttitle{Boulders on Ceres}

\begin{document}

\title{The brittle boulders of dwarf planet Ceres}

\correspondingauthor{Stefan Schr\"oder}
\email{stefanus.schroeder@dlr.de}

\author[0000-0003-0323-8324]{Stefan~E.~Schr\"oder}
\affiliation{Deutsches Zentrum f\"ur Luft- und Raumfahrt (DLR) \\
12489 Berlin, Germany}

\author[0000-0002-0859-4718]{Uri~Carsenty}
\affiliation{Deutsches Zentrum f\"ur Luft- und Raumfahrt (DLR) \\
12489 Berlin, Germany}

\author[0000-0002-1375-304X]{Ernst~Hauber}
\affiliation{Deutsches Zentrum f\"ur Luft- und Raumfahrt (DLR) \\
12489 Berlin, Germany}

\author[0000-0002-4213-8097]{Carol~A.~Raymond}
\affiliation{Jet Propulsion Laboratory (JPL), California Institute of Technology \\
91109 Pasadena, CA, U.S.A.}

\author[0000-0003-1639-8298]{Christopher~T.~Russell}
\affiliation{Institute of Geophysics and Planetary Physics (IGPP), University of California \\
90095-1567 Los Angeles, CA, U.S.A.}

\begin{abstract}

We mapped all boulders larger than 105~m on the surface of dwarf planet Ceres using images of the Dawn framing camera acquired in the Low Altitude Mapping Orbit ({\sc lamo}). We find that boulders on Ceres are more numerous towards high latitudes and have a maximum lifetime of $150 \pm 50$~Ma, based on crater counts. These characteristics are distinctly different from those of boulders on asteroid (4)~Vesta, an earlier target of Dawn, which implies that Ceres boulders are mechanically weaker. Clues to their properties can be found in the composition of Ceres' complex crust, which is rich in phyllosilicates and salts. As water ice is though to be present only meters below the surface, we suggest that boulders also harbor ice. Furthermore, the boulder size-frequency distribution is best fit by a Weibull distribution rather than the customary power law, just like for Vesta boulders. This finding is robust in light of possible types of size measurement error.

\end{abstract}



\section{Introduction} \label{sec:intro}

Boulders on planetary bodies bear information on past and present surface processes. In particular, the boulder properties and spatial distribution are related to the bulk properties of the parent body and the surface environmental conditions. On terrestrial planets, processes like impact cratering, volcanism, and mass wasting are typically responsible for the boulder formation. Degradation of boulders may result from processes like comminution by impacts and weathering, which on bodies with water and/or an atmosphere can include chemical weathering. Over the last decades, observations by spacecraft have revealed the existence of boulder populations on small airless Solar System bodies such as comets \citep{P15,PL16}, asteroids \citep{L96,T01,M08,K12,J15,M19,D19}, icy satellites \citep{P21}, and the protoplanet (4)~Vesta \citep{S20}. In the absence of an atmosphere and volatiles like water, there are only a few processes that can produce and destroy boulders. The most important formation mechanisms are the destruction of a parent body \citep{M20} and spallation during large impacts \citep{KK16}. The former is thought to be responsible for the boulder-dominated surfaces of rubble-pile asteroids \citep{F06,M19,D19}, whereas the latter process dominates on asteroids suspected to be more monolithic \citep{L96,T01,K12}. Destruction by small impacts \citep{B20} and thermal stress weathering \citep{D14,M17} are the most important degradational processes.

Dwarf planet (1)~Ceres maintains a position somewhat in between small bodies and the terrestrial planets, in the sense that it is a large, volatile-rich world, yet without atmosphere \citep{RR16}. As such, boulders on its surface may be affected by more processes than on small bodies, but by fewer than on the larger, more complex terrestrial planets. Here, we investigate the boulder population of Ceres and compare it with that of Vesta \citep{S20}. Both bodies were imaged by the same camera aboard the Dawn spacecraft \citep{RR11}. Such a comparison benefits from the fact that Vesta and Ceres have comparable distances to the Sun and very similar surface gravities \citep{B13}. So, any differences between the respective boulder populations may relate to compositional differences, with Ceres' crust harboring water ice, phyllosilicates, and salts \citep{A16,dS16,P17}, and Vesta's crust being basaltic \citep{dS12}. The limited spatial resolution of the global Dawn image data set restricts our study to clasts larger than 100~m, for which \citet{BR17} suggested the term megaclasts. But ``boulder'' has typically been used for clasts on small airless bodies irrespective of their size, and for consistency with the Vesta study we retain the term boulder.

Here, we study the global boulder population of Ceres using similar methods as those used for the Vesta boulder population \citep{S20}, as described in Sec.~\ref{sec:methods}. The results of our analysis are reported in Sec.~\ref{sec:results}. We searched all Dawn images acquired in the Low Altitude Mapping Orbit ({\sc lamo}) for boulders and determine general statistics of the global population related to boulder sizes and numbers (Sec.~\ref{sec:general_stats}). We determine the size-frequency distribution (SFD) of boulder populations of individual craters and that of the global population (Sec.~\ref{sec:size-freq-dist}). The boulder SFD is traditionally fit with a power law, but the Vesta boulders rather follow a Weibull distribution. We evaluate whether the same holds true for Ceres boulders. We investigate the spatial distribution of boulders in and around individual craters, as well as the distribution of craters with boulders across the globe (Sec.~\ref{sec:global_distribution}). Furthermore, we estimate the average boulder lifetime by comparing the boulder density around craters for which an age estimate is available, and assess the \citet{B15} prediction that meter-sized boulders on Ceres have the same lifetime as on Vesta (Sec.~\ref{sec:age}). In Sec.~\ref{sec:discussion} we discuss our results and the implications of the observed differences with the Vesta boulder population.

\section{Methods}
\label{sec:methods}

\subsection{Boulder mapping} \label{sec:mapping}

Boulders on Ceres can only be distinguished in framing camera images acquired in the Low Altitude Mapping Orbit ({\sc lamo}) at an altitude of around 400~km and lower orbits of the extended mission \citep{R07}. The framing camera is a narrow-angle camera with a field-of-view of $5.5^\circ \times 5.5^\circ$ \citep{S11}. {\sc lamo} coverage of the illuminated surface was near-complete for the camera's clear filter, but color imaging was sparse. The clear filter (F1) is a polychromatic filter with 98\% transmission in the 450 to 920~nm wavelength range \citep{S11}. {\sc lamo} images were acquired between 16 December 2015 and 27 August 2016 and have a typical scale of 35~m per pixel\footnote{When we talk about ``pixels'' in this paper, we are always referring to {\sc lamo} pixels.} \citep{R17}. The average scale of 59~{\sc lamo} images that we used in our analysis (at least one for each crater with boulders) is $35.8 \pm 1.3$~meter per pixel. The boulder-finding procedure was identical to that followed for Vesta boulders \citep{S20}. In summary, the second author browsed the entire data set of {\sc lamo} clear filter images and identified, measured, and mapped all boulders using the J-Ceres GIS program, which is a version of {\sc jmars}\footnote{\url{https://jmars.mars.asu.edu/}} \citep{CE09}, after which the first author reviewed the results for accuracy and completeness. Boulders were identified as positive relief features in projected images at a zoom level of 1024 pixels per degree. The {\sc lamo} resolution is about 230 pixels per degree at the equator, so this represents a zoom factor of about 4. Boulder size was determined using the J-Ceres crater measuring tool, which draws a circle around a boulder fitted to 3~points that are selected by the user on the visible boulder outline. The measurement uncertainty is about a single pixel. The limited accuracy of pointing information for {\sc lamo} images leads to mismatches between projected images. We used small craters inside and outside the crater as tie points to align the projected images to a Ceres background mosaic and relative to each other. All this leads to uncertainty in the location of boulders on the order of 500~m. We are confident that we could reliably identify boulders with a size of at least 3~pixels (105~m), although a criterion of 4~pixels (140~m) is more likely to ensure that mapping is complete \citep{S20,P21}. We did not distinguish between boulders located either inside or outside the crater rim, a choice that we justify in Sec.~\ref{sec:global_distribution}.

\begin{figure}
	\centering
	\includegraphics[width=8cm,angle=0]{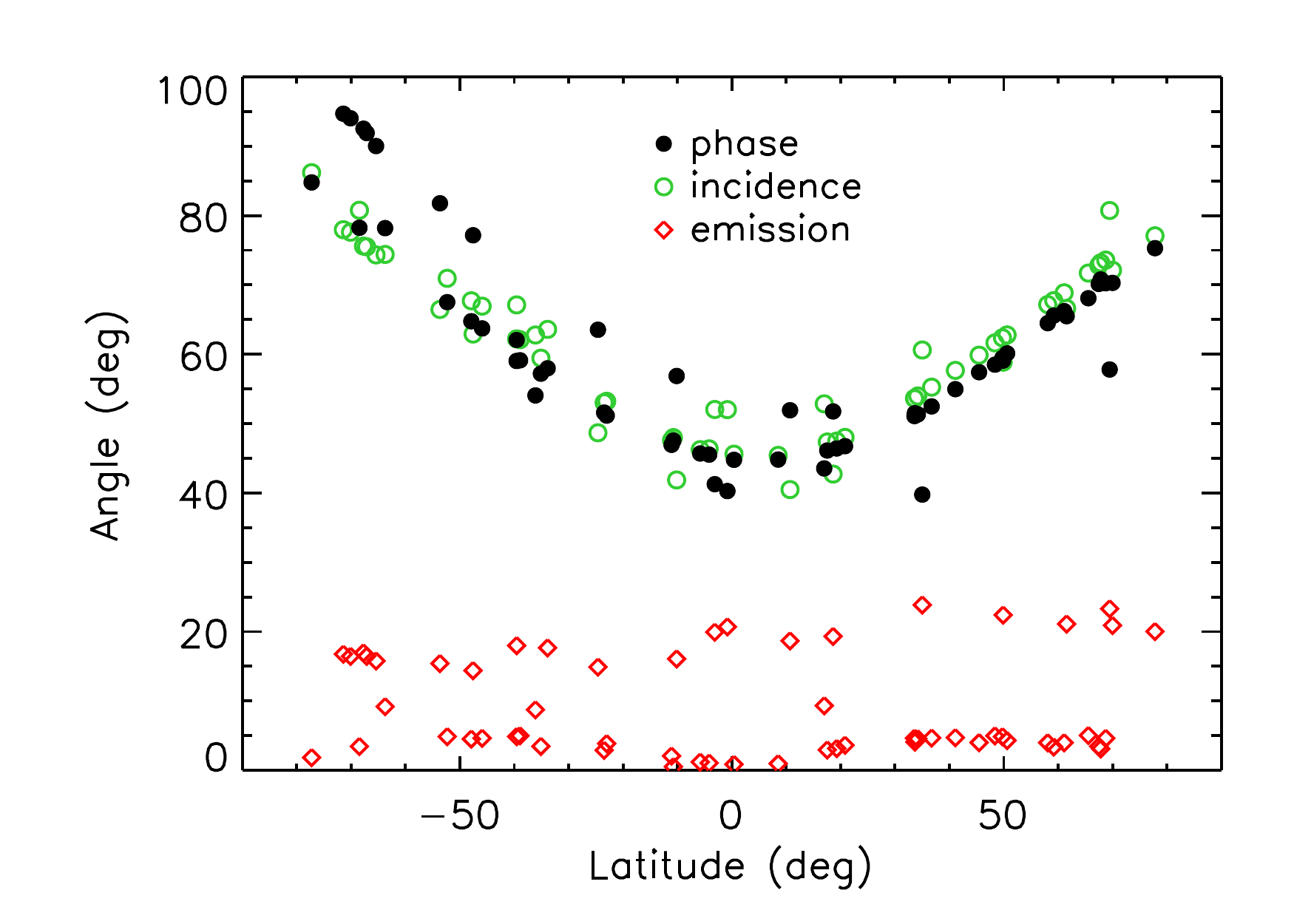}
	\caption{Boulder viewing conditions: Photometric angles at the center of selected {\sc lamo} images for an ellipsoid Ceres.}
	\label{fig:viewing_angles}
\end{figure}

\begin{figure}
	\centering
	\includegraphics[width=\textwidth,angle=0]{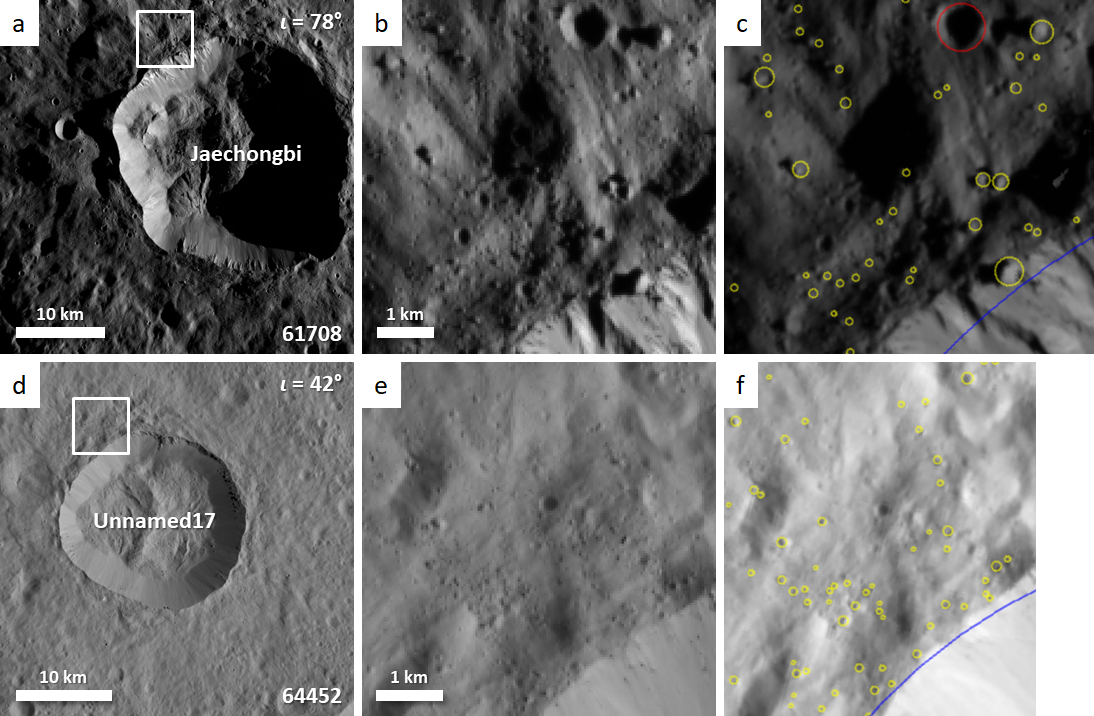}
	\caption{How does the solar incidence angle ($\iota$) affect the visibility of boulders? {\bf a}-{\bf c}.~Jacheongbi crater at $\iota = 78^\circ$. {\bf d}-{\bf f}. Crater Unnamed17 at $\iota = 42^\circ$. {\bf a} and {\bf d} show the full {\sc lamo} image with the outlined area enlarged in {\bf b} and {\bf e}. {\bf c} and {\bf f} are map-projected versions of the areas in b and e and show the boulders as mapped in J-Ceres (yellow circles are boulders, red circle is reference crater, and blue curve denotes crater rim). The incidence angle was calculated at the image center for an ellipsoid surface. The FC2 image number is indicated.}
	\label{fig:mapping}
\end{figure}

The illumination conditions at the time of imaging affect the visibility of a boulder, mainly by the strength of its shadow. The photometric angles at the center of the {\sc lamo} images, calculated for an ellipsoid Ceres, are plotted as a function of latitude $L$ in Fig.~\ref{fig:viewing_angles}. We see that the illumination conditions for imaging, and thereby boulder visibility, systematically changed according to latitude. The spacecraft looked at nadir most of the time (low emission angle), but the incidence and phase angle increased with $L$. If we distinguish three latitudinal zones as ``low'' ($|L| < 30^\circ$), ``mid'' ($30^\circ < |L| < 60^\circ$), and ``high'' ($|L| > 60^\circ$), the average incidence angle at the image center is $\iota = 48^\circ \pm 4^\circ$ for low latitudes, $\iota = 62^\circ \pm 5^\circ$ for mid-latitudes, and $\iota = 75^\circ \pm 5^\circ$ for high latitudes. For a spherical boulder that is half-buried in a plane surface, the maximum length of the shadow is $l = r (\cos^{-1} \iota - \cos \iota)$, with boulder radius $r$ and incidence angle $\iota$. A boulder at low latitudes will cast the shortest shadow, with $l = 0.83 r$. A boulder with a diameter of 3 and 4~pixels will cast a shadow of 1.2 and 1.6~pixels, respectively. Especially for the larger diameter, the shadow is long enough to be well visible. Thus, although boulders will be more easily recognized in high-latitude images, we are confident that all boulders with a diameter of 4~pixels can be recognized in low-latitude images. Nevertheless, it is likely that we missed boulders with a 3-pixel diameter in low-latitude images due to their short shadows. In Fig.~\ref{fig:mapping} we investigate how the increase of $\iota$ with latitude affects our mapping. The figure shows {\sc lamo} images of two craters with abundant boulders: high-latitude crater Jacheongbi ($69^\circ$S, $\iota = 78^\circ$) and low-latitude crater Unnamed17 ($10^\circ$S, $\iota = 42^\circ$). Both craters appear fresh, with boulders that are easily recognized by the shadows they cast on their ejecta blankets. At {\sc lamo} resolution, the blankets appear equally smooth at either incidence angle. The rightmost panels show the distribution of boulders as mapped using J-Ceres. The stronger shadows of the Jacheongbi boulders did not lead us to recognize them in higher numbers compared to Unnamed17, which confirms that differences in visibility may only be consequential at the most extreme incidence angles \citep{W05}. The figure also shows that half of Jacheongbi's interior is in shadow. Boulders are abundant on the sunlit half of the crater floor, which suggests that boulder numbers in high-latitude craters are severely underestimated, with potentially important consequences for the SFD.

\subsection{Power law SFD} \label{sec:power_law}

Various practices for displaying boulder SFDs seen in the literature were discussed by \citet{S20}. In this paper, we display the SFD in both cumulative and differential format, using the incremental (binned, histogram) version for the latter \citep{C93}. The cumulative distribution of boulders on Solar System bodies is often assumed to follow a power law, for which the number of boulders with a size larger than $d$ is:
\begin{equation}\label{eq:ML}
N(>d) = N_{\rm tot} \left( \frac{d}{d_{\rm min}} \right)^\alpha,
\end{equation}
with $\alpha < 0$ the power law exponent and $N_{\rm tot}$ the total number of boulders larger than $d_{\rm min}$. The exponent of a cumulative distribution of a quantity that follows a power law is identical to that of the associated incremental differential distribution with a constant bin size on a logarithmic scale, if the logarithmic bins are chosen wide enough \citep{H69,C93}. The power law exponent is best estimated from the SFD by means of the maximum likelihood (ML) method \citep{N05,C09}. The ML power law exponent ($\alpha < 0$) is estimated directly from the boulder size measurements as
\begin{equation}\label{eq:ML_exp}
\hat{\alpha} = -N \left( \sum_{i=1}^{N} \ln \frac{d_i}{d_{\rm min}} \right)^{-1},
\end{equation}
with $d_i$ the size of boulder $i$ and $N$ the total number of boulders with a size larger than $d_{\rm min}$. The standard error of $\hat{\alpha}$ is
\begin{equation}\label{eq:ML_err}
\sigma = -\hat{\alpha} / \sqrt{N}
\end{equation}
plus higher order terms, which we ignore. The estimator in Eq.~\ref{eq:ML_exp} is unbiased only for sufficiently large sample size. \citet{C09} also provide details to a statistical test that evaluates whether a power law is an appropriate model for the data\footnote{\url{http://tuvalu.santafe.edu/~aaronc/powerlaws/}}. The test randomly generates a large number of synthetic data sets according to the best-fit power law model (specified by $\hat{\alpha}$ and $d_{\rm min}$), and calculates for each the Kolmogorov-Smirnov statistic, which is a measure of how well the synthetic data agree with the model. A $p$-value, defined as the fraction of synthetic data sets that have a larger statistic than the real data set, quantifies how well the power law performs. The authors adopt $p < 0.1$ to mean rejection of the power law model.

We fitted power laws to the global boulder population, but also to the populations associated with individual craters to investigate possible variations of the exponent over the surface. To evaluate whether any such variations found are meaningful, we simulate the Ceres boulder population on the basis of the power law, using the observed population sizes of individual craters as input. For all craters we adopt the same exponent, namely that of the power law that best fits the global population. The continuous power law probability distribution is also known as the Pareto distribution \citep{N05}. To simulate a size distribution of boulders associated with impact craters we draw a random variate $U$ from a uniform distribution on $(0, 1)$ using the {\sc randomu} routine in IDL with an undefined seed. Then the boulder diameter
\begin{equation}\label{eq:Pareto}
d = d_{\rm min} U^{1/\alpha}
\end{equation}
follows a Pareto distribution, with $\alpha$ the power law index (associated with the cumulative distribution function, with $\alpha < 0$). We adopted a minimum boulder diameter of $d_{\rm min} = 140$~m (4~pixels). We simulated populations for all craters and estimated the power law exponent for each using the ML method. We then compared the resulting distribution of exponents with the observed one.

\begin{figure}
	\centering
	\includegraphics[width=\textwidth,angle=0]{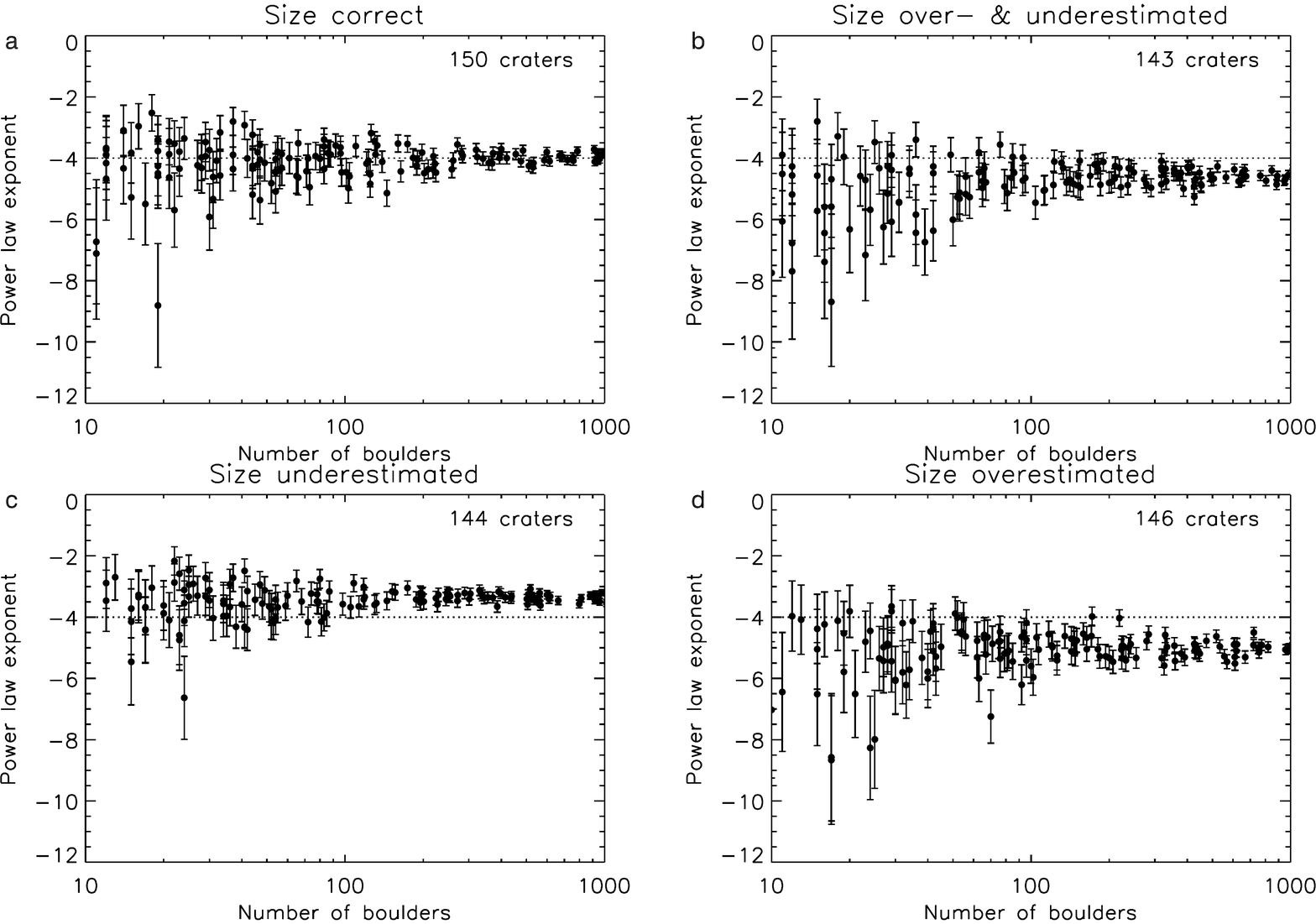}
	\caption{Investigating the effect of measurement error on the power law exponent estimated for simulated craters with boulder populations that have an exponent of $-4$ (dotted line). The pixel size is 35~m, and the fit was performed on boulders ``measured'' larger than 4~pixels (140~m). {\bf a}.~Boulder sizes measured correctly. {\bf b}.~Sizes either correctly measured, overestimated by 1~pixel, or underestimated by 1~pixel with equal probability. {\bf c}.~Sizes underestimated by 1~pixel. {\bf d}.~Sizes overestimated by 1~pixel. The number of data points (craters) in the plot is indicated.}
	\label{fig:measunc_crater}
\end{figure}

The power law exponent estimated according to Eq.~\ref{eq:ML_exp} is biased for low boulder numbers \citep{C09}, which was illustrated for simulated boulder populations by \citet{S20}. Another potential source of bias is measurement error of boulder size. While the boulders described in this paper are large in absolute terms, they typically measure only a few pixels across, and measurement errors on the order of a pixel can be expected. Here we investigate the consequences of measurement error using boulder populations simulated according to Eq.~\ref{eq:Pareto}. First, we consider a group of ``craters'', each with a simulated boulder population of different size. The SFD of all boulder populations follows a power law with exponent $-4$, with a minimum boulder size of 40~m. We modified the boulder sizes according to three different definitions of measurement error, where we distinguish between systematic and random errors. Measurement errors are sized one pixel of 35~m, equal to the spatial resolution of the Ceres {\sc lamo} images. We estimated the power law exponent for each crater, including only boulders larger than 4~pixels (140~m) in the fit. Note that we can only assess how many boulders have met this requirement after performing the simulation. Figure~\ref{fig:measunc_crater} shows the results. In (a), the boulder sizes are all measured correctly, and the power law exponent is retrieved reliably for craters with large boulder numbers. The negative bias at small boulder numbers inherent in Eq.~\ref{eq:ML_exp} can be recognized clearly. In (b), boulder sizes are affected by measurement errors with a random character: Sizes are either decreased by 1~pixel (size overestimated), increased by 1~pixel (size underestimated), or left unchanged, each with equal probability of $1/3$. Figures (c) and (d) explore the consequences of systematic measurement errors, by either under- or overestimating all boulder sizes by 1~pixel. Systematic errors can be introduced by the method of measuring. Figure~\ref{fig:measunc_crater} shows that all types of measurement error lead to biased power law exponents. Underestimating the boulder sizes increases the exponent by a little less than unity (shallower power law), while overestimating the sizes decreases the exponent by unity (steeper power law). The bias is stronger for overestimating the sizes than for underestimating, which cause the random measurement errors in (b) decrease the exponent by a little less than unity (steeper power law). Another consequence of random measurement errors is that the exponent converges only at larger boulder numbers than without errors, which is not reflected in the formal uncertainty of the exponent (Eq.~\ref{eq:ML_err}).

\begin{figure}
	\centering
	\includegraphics[width=\textwidth,angle=0]{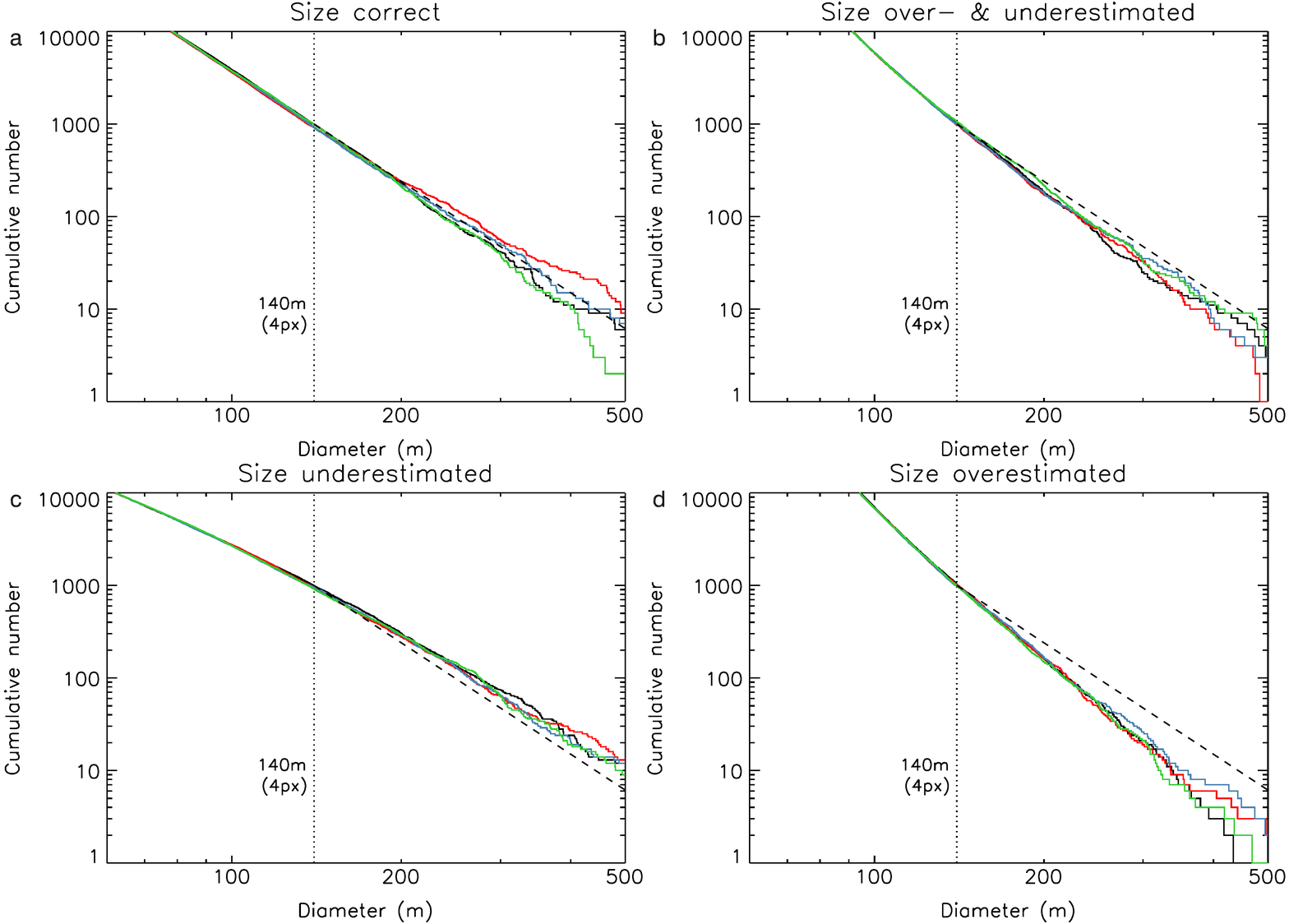}
	\caption{Investigating the effect of measurement error on the shape of the cumulative distribution when the true power law exponent is $-4$. The pixel size is 35~m, and the adopted power law is only shown for sizes ``measured'' larger than 4~pixels (140~m, dashed line). The four curves of different color represent repeated simulations of the same population. {\bf a}.~Boulder sizes measured correctly. {\bf b}.~Sizes either correctly measured, overestimated by 1~pixel, or underestimated by 1~pixel with equal probability. {\bf c}.~Sizes underestimated by 1~pixel. {\bf d}.~Sizes overestimated by 1~pixel. The population sizes for the four different scenarios were adjusted to achieve a cumulative number of boulders of around 1000 at 140~m.}
	\label{fig:measunc_cumul}
\end{figure}

Another aspect of this problem is the shape of the cumulative SFD. Figure~\ref{fig:measunc_cumul} investigates how the shape changes for the same three cases of random and systematic measurement errors. For each case we generated four populations of the same size, again with a power law exponent of $-4$. In (a), the boulder sizes are all measured correctly, and the SFD follows the straight line of the power law up to a diameter of about 200~m. Beyond this size, the simulated curves diverge considerably due to chance. In (b), the boulder sizes are affected by measurement errors with a random character, which steepens the SFD slightly. In (c), boulder sizes are systematically underestimated, which makes the SFD shallower and introduces a slightly convex curvature. In (d), boulder sizes are systematically overestimated, which steepens the SFD considerably and introduces a slightly concave curvature. Our simulations demonstrate that bias due to measurement error is unavoidable when typical boulder sizes are on the order of a few image pixels. The results in Figs.~\ref{fig:measunc_crater} and \ref{fig:measunc_cumul} may allow us to identify or predict such bias for the Ceres boulder population.

\subsection{Weibull SFD}

The SFD of the Vesta boulder population is better described by a Weibull distribution than a power law \citep{S20}. We investigate whether the same holds true for the Ceres boulder population. The Weibull distribution was initially derived empirically, and is often used to describe the particle distribution resulting from grinding experiments \citep{RR33}. Where the power law follows naturally from a single-event fragmentation that leads to a branching tree of cracks that have a fractal character, the Weibull distribution results from sequential fragmentation \citep{BW95}. Because we only include boulders larger than a certain size in the fit, we employ a left-truncated Weibull distribution with the cumulative form \citep{W89}:
\begin{equation}\label{eq:Weibull_min_size}
N(>d) = N \exp [ - \alpha ( d_i^\beta - d_{\rm min}^\beta ) ],
\end{equation}
where $N$ is the number of boulders larger than $d_{\rm min}$. We estimate the Weibull parameters $\alpha$ and $\beta = 3(\gamma + 1)$ from the boulder sizes $d_i > d_{\rm min}$ using the ML method. To maximize the log-likelihood function, these two equations must be satisfied:
\begin{eqnarray}\label{eq:Weibull_estimation}
	\alpha = \frac{N}{\sum (d_i^\beta - d_{\rm min}^\beta)} \\
	\frac{N}{\beta} + \sum \ln d_i - N \frac{\sum (d_i^\beta \ln d_i - d_{\rm min}^\beta \ln d_{\rm min})}{\sum (d_i^\beta - d_{\rm min}^\beta)}= 0.
\end{eqnarray}
We find $\hat{\beta}$ from a simple grid search, and $\hat{\alpha}$ by inserting $\hat{\beta}$.

\section{Results}
\label{sec:results}

\subsection{General statistics}
\label{sec:general_stats}

\begin{figure}
	\centering
	\includegraphics[width=\textwidth,angle=0]{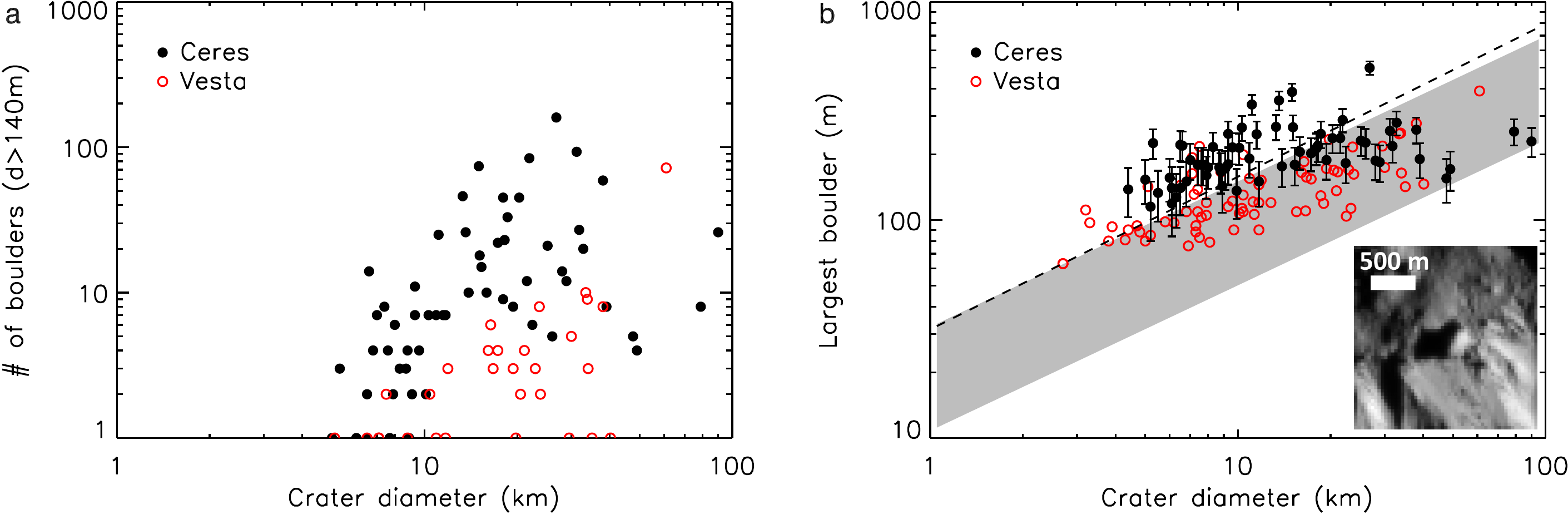}
	\caption{General statistics of boulders associated with craters on Ceres (this paper) and Vesta \citep{S20}. {\bf a}.~Number of boulders larger than 140~m as a function of crater diameter. {\bf b}.~Diameter of the largest boulder as a function of crater diameter. The uncertainty of the Ceres boulder size derives from a measurement error of 1~pixel (Vesta boulder error bars are omitted for clarity). The empirical range given by \citet{M71} for selected lunar and terrestrial craters is shown in gray. The dashed line is the relation given by \citet{L96}. The inset shows the largest boulder identified on Ceres, a 500~m sized block on the rim of Jacheongbi crater.}
	\label{fig:general_stats}
\end{figure}

We identified a total of 4423 boulders on the surface of Ceres with a diameter larger than 3~image pixels (105~m), of which 1092 were larger than 4~pixels (140~m). All boulders are associated with impact craters. The details of all craters with at least one boulder larger than 4~pixels ($n = 58$) are listed in Table~\ref{tab:craters}. First, we summarize some general statistics of the global boulder population. Figure~\ref{fig:general_stats}a shows a (weak) correlation between the number of boulders per crater and crater size. The largest craters in the sample have a number of boulders that is much smaller than expected. For example, Occator is the largest crater with a diameter of 90~km, and it has 26 boulders larger than 4~pixels. From the general trend in the figure, we can expect to find a number at least an order of magnitude larger for a crater of its size. But many of its boulders may have been destroyed or hidden from view by the large-scale flows that are present in and around the crater \citep{SS19}. In fact, three out of the largest four craters in our sample show evidence of such flows: Azacca, Ikapati, and Occator \citep{B18,K18,SS19}. Boulders around these craters are also difficult to distinguish from features resulting from partly submerged topography. The fourth crater in this group, Gaue, is very old \citep{S16,P18}. Age must affect the number of boulders, as they are destroyed over time. The figure also shows the number of boulders associated with craters on Vesta \citep{S20}, where we adopted the same minimum size (140~m) as for the Ceres boulders. Clearly, the number of boulders produced on average by impacts on Ceres is larger than on Vesta for craters of the same size.

We investigate the relation between the size of the largest boulder ($L$) and size of the crater ($D$) in Fig.~\ref{fig:general_stats}b. Being a single measurement, the largest boulder size is a poor statistic, but is nevertheless often used to characterize boulder populations \citep{T01,K12,J15,S18}. We compare the Ceres distribution with the relation provided by \citet{L96} for craters formed in rocky targets ($L = 0.25 D^{0.7}$ with $L$ and $D$ in m) and with the empirical range established by \citet{M71} for a selection of lunar and terrestrial craters ($L = 0.01^{1/3} K D^{2/3}$ with $K$ ranging from 0.5 to 1.5). The former relation represents more or less the upper limit of the latter range. The largest boulders on Ceres do not agree well with either relation from the literature, with sizes that are almost independent of the size of the craters. Again, it is mostly the largest craters that break the trend, probably because the aforementioned flow features and old age. The largest boulder we found on Ceres is a 500~m large block on the rim of Jacheongbi (figure inset), a relatively large crater (27~km). The figure also shows the largest boulders of Vesta craters. While the Vesta data also do not perfectly agree with either relation from the literature, most fall within the range of \citet{M71}. On average, the largest boulders on Vesta are somewhat smaller than those on Ceres.

\subsection{Size-frequency distribution}
\label{sec:size-freq-dist}

\begin{figure}
	\centering
	\includegraphics[width=\textwidth,angle=0]{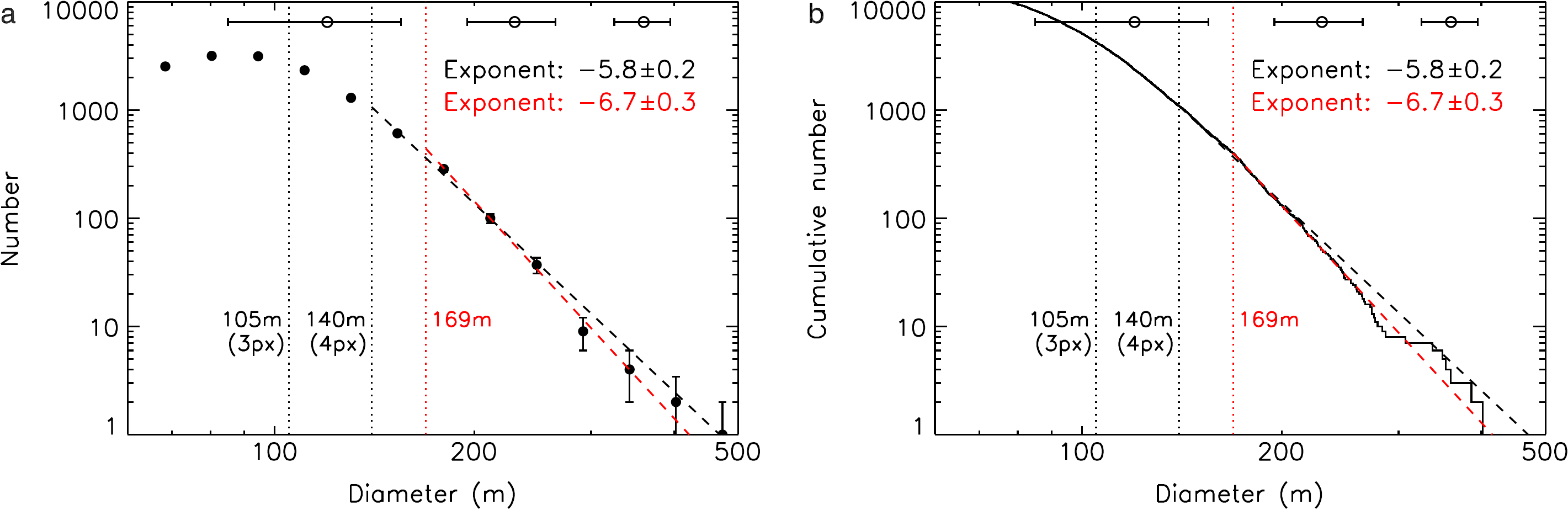}
	\caption{The SFD of all boulders identified on Ceres, displayed both in differential ({\bf a}) and cumulative ({\bf b}) format. Different size limits are indicated by vertical (dotted) lines. The dashed lines are best-fit power laws using the ML method, with exponent indicated: The black dashed line has $d_{\rm min} = 140$~m (4~pixels), whereas the red dashed line has $d_{\rm min} = 169$~m, as estimated by the ML algorithm. The error bars at the top indicate the uncertainty in boulder size at different diameters due to a 1~pixel measurement error.}
	\label{fig:global_ML}
\end{figure}

We aggregate all boulders counted on the surface of Ceres to find the cumulative power law exponent of the global boulder population. We note that the resulting global SFD is biased, as the boulder populations of the largest craters were almost certainly decimated by large scale flows (see Sec.~\ref{sec:general_stats}). Figure~\ref{fig:global_ML} shows the SFD, both in cumulative and differential representation. At the top of the differential plot we show the uncertainty in size resulting from a 1~pixel measurement error. We chose a logarithmic bin size of 0.07 with the boulder size in meters, ensuring that the size is on the order of the measurement error at the larger end of the scale. As the error is larger than the bin size at the smaller end of the scale, we can expect boulders to end up in adjacent bins merely by chance. We recognize the characteristic roll-over of the distributions towards smaller diameters, caused by the limited spatial resolution and the measurement error. We fit two power laws to the data with the ML method, one with the minimum boulder size ($d_{\rm min}$) fixed, and the other with $d_{\rm min}$ estimated by the ML algorithm. When fixing $d_{\rm min}$ to 4~pixels (140~m), we find a power law exponent of $\alpha = -5.8 \pm 0.2$ ($n = 1092$, black dashed lines in Fig.~\ref{fig:global_ML}). By extrapolating the power law to smaller diameters we find that the number of boulders with a diameter around 3~pixels may be severely underestimated; the observed number in the bin closest to the 3~pixel limit is 2328, but the extrapolated, expected number is about 4000. The counts for boulders larger than 4~pixels are probably close to complete. We note that the counts at the largest diameters do not match well with the power law, both in the cumulative and differential representation. The statistical test provided by \citet{C09} confirms  that this power law is not a good model for the data ($p = 0$). When we let the ML algorithm itself choose the minimum boulder size, we find a larger $d_{\rm min} = 169$~m and a steeper power law with $\alpha = -6.7 \pm 0.3$ ($n = 400$, red dashed lines in Fig.~\ref{fig:global_ML}). The statistical test indicates that this power law is a good model for the data ($p = 0.37$).

\begin{figure}
	\centering
	\includegraphics[width=\textwidth,angle=0]{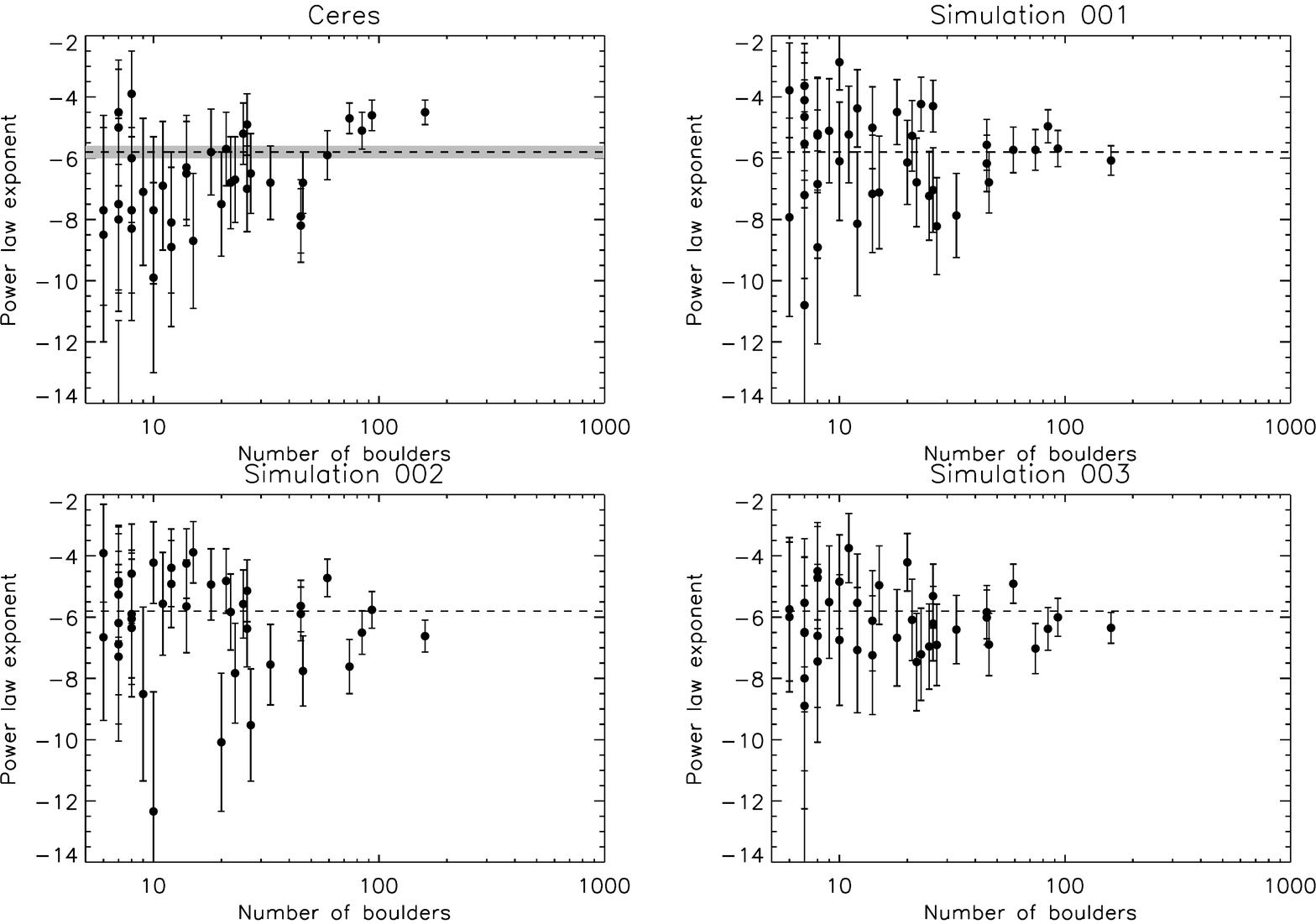}
	\caption{Power law exponents for all craters with a population of at least 6 boulders larger than 4~pixels ($n = 39$). The observed exponents were derived by fitting a power law to the data of each crater. The best fit power law index for the observed global boulder distribution is $\alpha = -5.8 \pm 0.2$ (dashed line with gray confidence interval). The crater with the largest number of boulders (160) is Jacheongbi. We compare the observations to three simulations. The simulated exponents were derived by fitting randomly generated boulder distributions, assuming a Pareto distribution with $\alpha = -5.8$ (dashed line), using the number of boulders in the population of each crater as input.}
	\label{fig:accuracy}
\end{figure}

We also estimated the power law exponent for individual craters that have at least 6~boulders larger than 4~pixels (Table~\ref{tab:craters}), and plot these as a function of number of boulders in the population in Fig.~\ref{fig:accuracy}. The figure also includes three simulations of the power law exponent distribution of individual craters. The simulation uses the observed population sizes and adopt the best-fit power law exponent for the global boulder population ($\alpha = -5.8$). The observations and simulations agree in showing a large scatter in the exponents for smaller population sizes and their expected negative bias \citep{C09,S20}. The degree of scatter is similar for both simulated and observed data, indicating that the observed variety in power law exponents is merely due to differences in population size and not some physical property of boulders or craters. However, the observed exponents of craters with small boulder populations are typically more negative than in the simulations. Additionally, the power law exponent of the crater with the largest number of boulders, Jacheongbi, is further from $-5.8$ than that of the simulated ``Jacheongbi's''. It is almost as if the observed distribution of exponents is skewed with respect to the simulated distribution. This suggests that the power law model does not correctly describe the boulder SFD.

\begin{figure}
	\centering
	\includegraphics[width=6cm,angle=0]{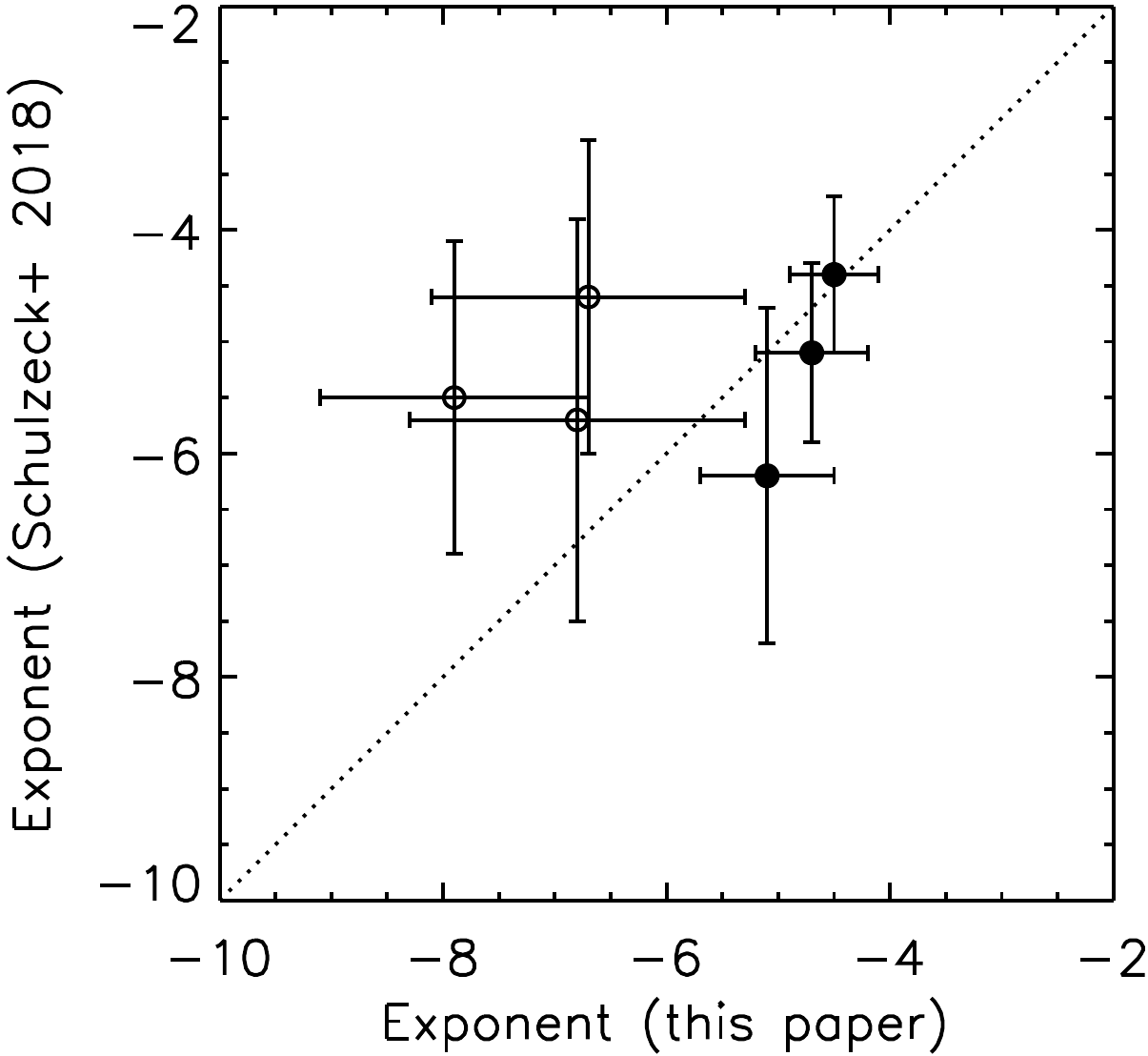}
	\caption{Comparison of the power law exponents for the craters Juling, Jacheongbi, Nunghui, Ratumaibulu, Unnamed11, and Unnamed17 as determined in this paper and by \citet{S18}. Filled symbols refer to exponents that are more reliable, being associated with populations of more than 70~boulders.}
	\label{fig:exponent_comparison}
\end{figure}

\citet{S18} independently counted boulders around several Ceres craters and fitted power laws to the SFD, using ML to estimate both the exponent and the minimum diameter. We compare their exponents with ours in Fig.~\ref{fig:exponent_comparison}. The exponents agree within the error bars for the craters with more than 70~boulders (Jacheongbi, Nunghui, and Unnamed11). The exponents for the other three craters (Juling, Ratumaibulu, and Unnamed17) agree less well, which is not surprising given the small size of their boulder populations. The exponents for the crater with the largest number of boulders (Jacheongbi with 160~boulders) match closely ($-4.5 \pm 0.4$ versus $-4.4 \pm 0.7$). This suggests that our counts are consistent with those of \citet{S18}. In Sec.~\ref{sec:power_law} we uncovered evidence for bias resulting from measurement errors in the boulder sizes, which were probably similar in magnitude for \citet{S18}. Can this bias be responsible for the unusual steepness and the convex shape of the SFD of the global boulder population? Given that the boulder size measurements were almost certainly subject to errors of at least one image pixel, the ``true'' SFD is probably less steep (see Figs.~\ref{fig:measunc_crater} and \ref{fig:measunc_cumul}). The power law exponent may be smaller by about unity, but with a value somewhere between $-4.8$ and $-5.7$, the SFD would still be unusually steep. Measurement errors also affect the shape of the SFD (Fig.~\ref{fig:measunc_cumul}). Random errors would actually lead to a slightly more concave shape of the SFD. Only systematically underestimating the boulder sizes would lead to a convex SFD, but this would also tend to make it shallower. Thus, we cannot attribute the downturn of the SFD towards large sizes to measurement error, and it must be an intrinsic property of the Ceres boulder population.

\begin{figure}
	\centering
	\includegraphics[width=12cm,angle=0]{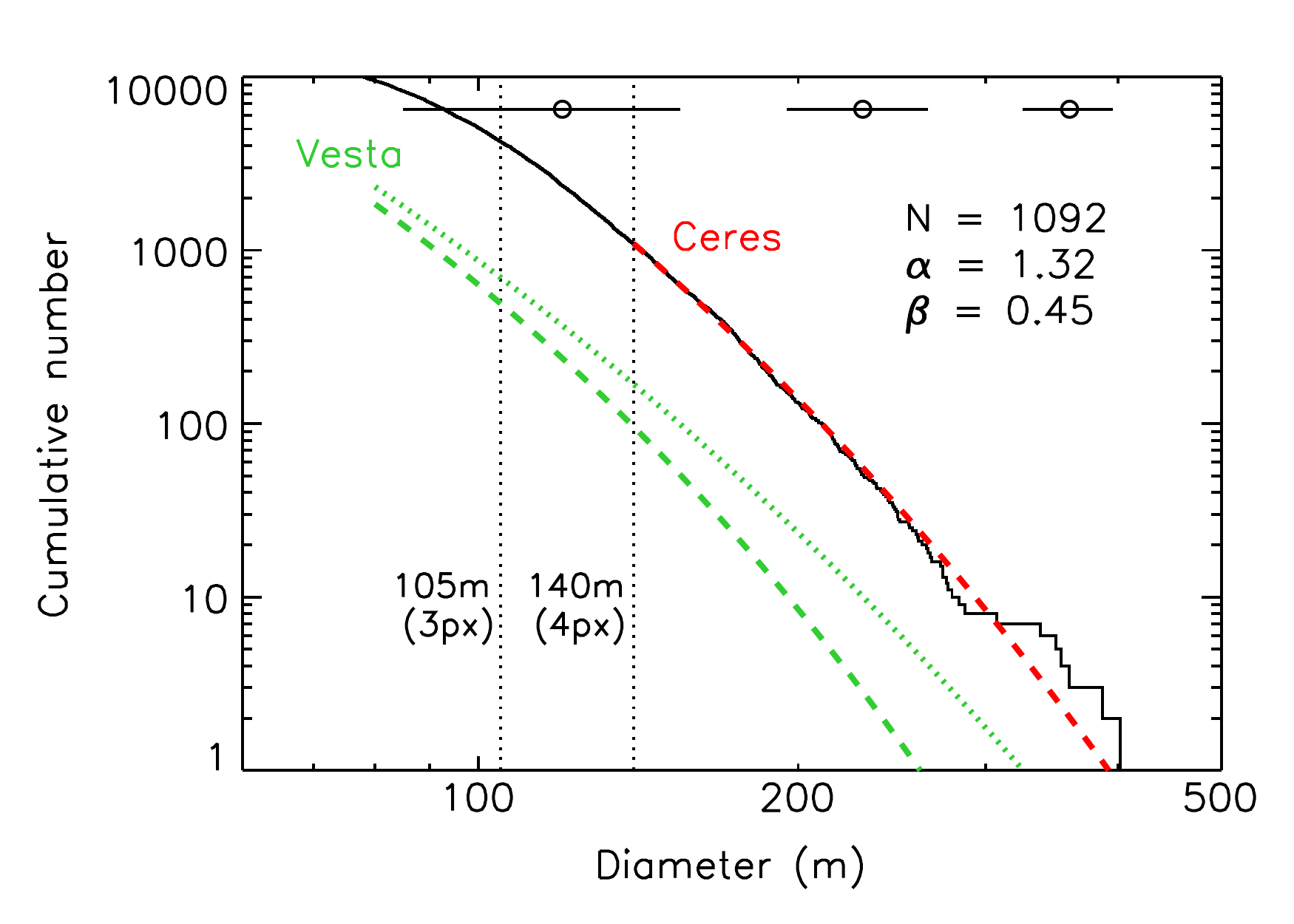}
	\caption{Cumulative size-frequency distribution for Ceres boulders larger than 4~pixels (black curve). The parameters of the best fit left-truncated Weibull distribution (Eq.~\ref{eq:Weibull_min_size}, red curve) are listed. The error bars at the top indicate the uncertainty in boulder size at different diameters assuming a 1~pixel measurement error. Also shown are Weibull fits for the Vesta global boulder population (green curves), including (dotted) and excluding (dashed) boulders from the largest crater, Marcia \citep{S20}}
	\label{fig:Weibull}
\end{figure}

We conclude that the power law is not a good model for the SFD of boulders larger than 4~pixels. The ML algorithm was able to find an acceptable power law for a larger minimum size ($d_{\rm min} = 169$~m), but there is no reason to exclude the well-resolved boulders larger than 4~pixels but smaller than 169~m (more than half of the total) from the fit. This is the same situation as for the Vesta boulder SFD, for which the Weibull distribution proved to be a better model than the power law \citep{S20}. The best-fit Weibull distribution for the Ceres global boulder population has $N = 1092$, $\alpha = 1.32$, and $\beta = 0.45$ (Fig.~\ref{fig:Weibull}). The fractal dimension $D_{\rm f} = 3 - \beta$ for the cracks in the rock is 2.5. The Weibull distribution fits the SFD better than the power law, and, contrary to the latter, it does not predict that the number of boulders with a size of 3~pixels is massively underestimated. Figure~\ref{fig:Weibull} also shows Weibull distributions for the Vesta global boulder population \citep{S20}. There are two best-fit curves, one including and the other excluding boulders of Marcia, the largest crater on Vesta. Just like the largest Ceres craters, Marcia shows evidence of flows, which may have destroyed many of its boulders. As Vesta is smaller than Ceres, its Weibull distributions plot below the Ceres distribution. Given the uncertainty surrounding the boulder populations of the largest craters on both worlds, a detailed comparison of the Weibull parameters provides little insight.

\subsection{Spatial distribution}
\label{sec:global_distribution}

\begin{figure}
	\centering
	\includegraphics[width=\textwidth,angle=0]{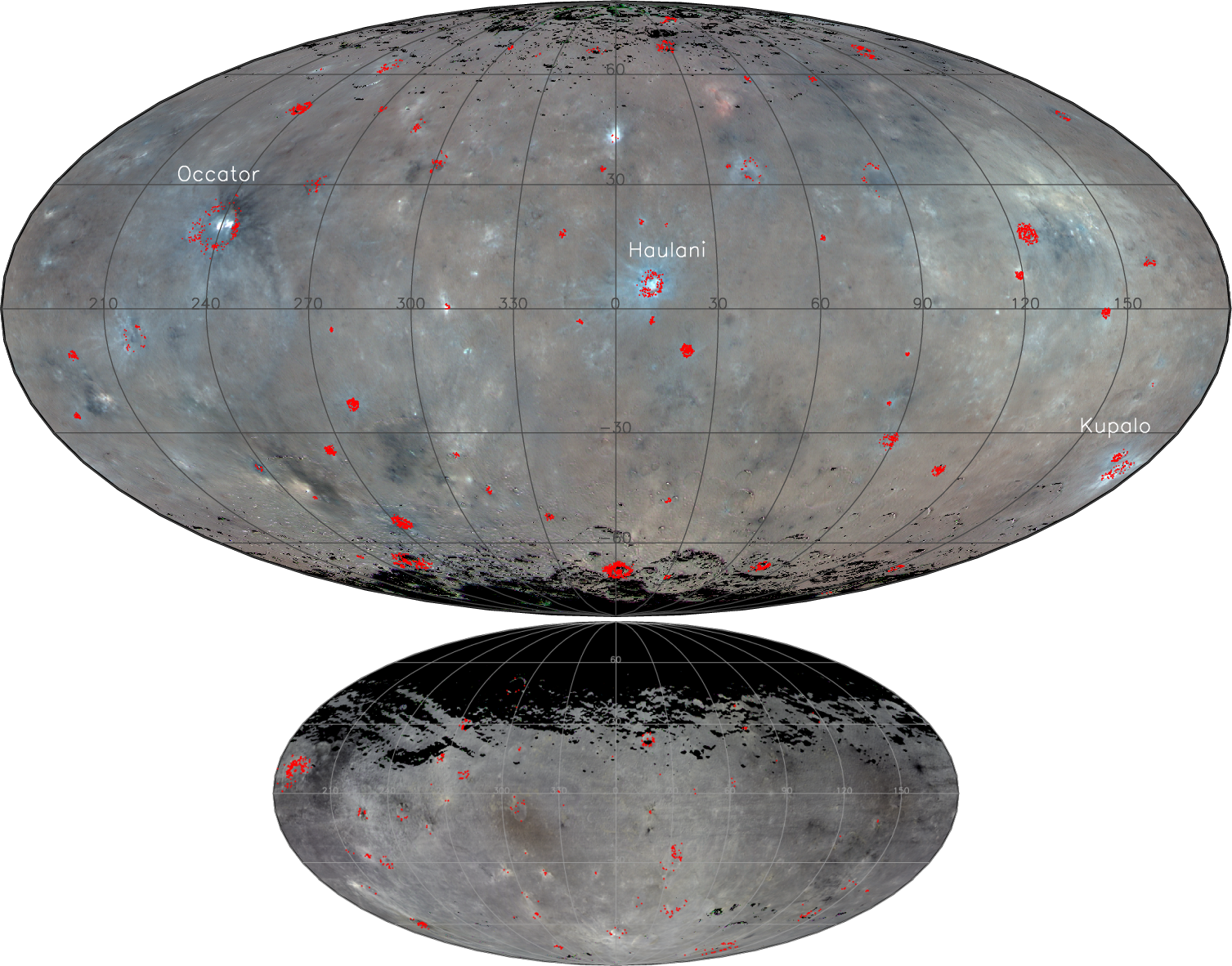}
	\caption{Distribution of boulders in Mollweide maps of Ceres ({\bf top}) and Vesta ({\bf bottom}), shown on the same scale. Shown in red are the locations of boulders with a size of at least 105~m (3~Ceres {\sc lamo} pixels) on photometrically corrected color maps. The Ceres map has filters centered at 965~nm, 555~nm, and 438~nm in the RGB color channels \citep{S17}, and the Vesta map has filters centered at 650~nm, 555~nm, and 438~nm in the RGB channels \citep{S13a}. The center of the maps is at (lat, lon) $= (0^\circ, 0^\circ)$.}
	\label{fig:global}
\end{figure}

In Fig.~\ref{fig:global}, we plot the distribution of all boulders with a size of at least 3 pixels (105~m) on a color composite map of Ceres. In the same figure and on the same scale, we also show the distribution of boulders larger than 105~m on Vesta using counts from \citet{S20}. With a surface area 3.2 times that of Vesta, the Ceres map is much larger. On both bodies, boulders are confined to craters. On Vesta, boulders are mostly absent from a large area of low albedo that appears to be enriched with carbonaceous chondrite material \citep{D16,S20}. The situation is different on Ceres. Several distinctly blue craters (Occator, Haulani, Kupalo) have boulders, which is consistent with blue being a marker of youth \citep{SK14}. Large areas are devoid of boulders, especially at lower latitudes, but these do not stand out in terms of color or albedo like on Vesta. Even though the poles were partly in shadow during {\sc lamo}, we find many boulders there.

\begin{figure}
	\centering
	\includegraphics[width=\textwidth,angle=0]{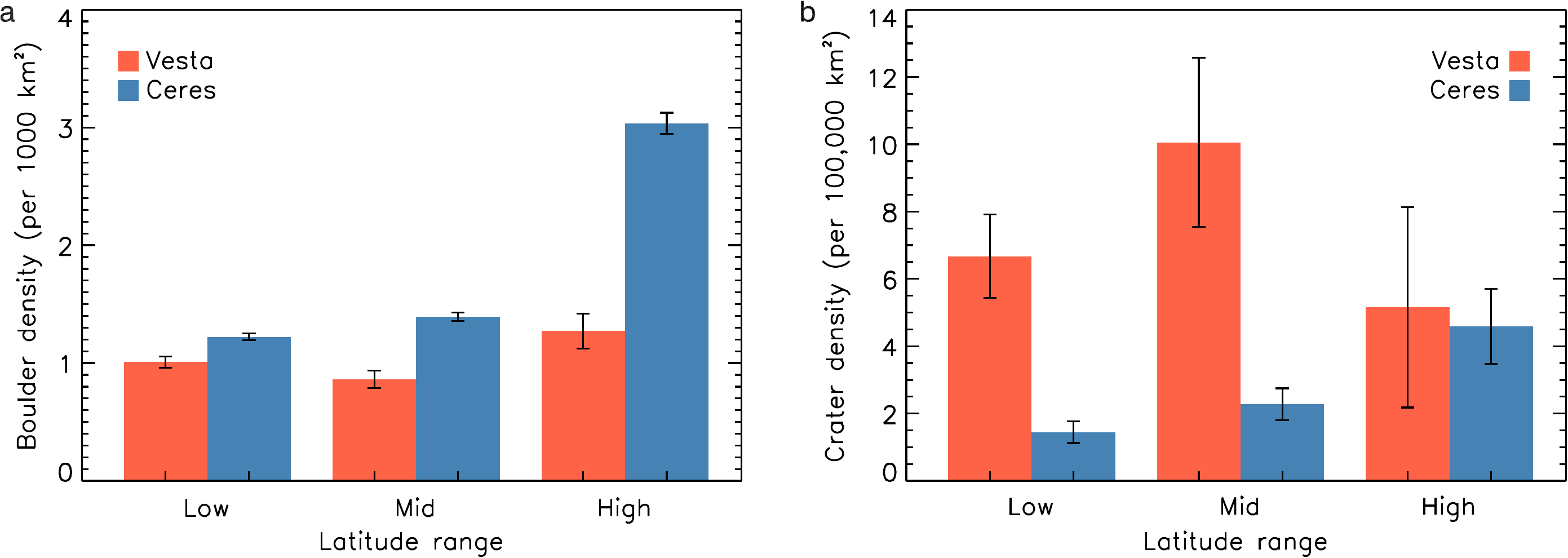}
	\caption{The distribution of boulders larger than 105~m on Vesta and Ceres along latitude $L$. We aggregate the boulders in three latitude ranges: ``low'' ($|L| < 30^\circ$), ``mid'' ($30^\circ < |L| < 60^\circ$), and ``high'' ($|L| > 60^\circ$). {\bf a}.~Number density of boulders. {\bf b}.~Number density of craters with at least one boulder larger than 105~m. The error bars derive from Poisson statistics.}
	\label{fig:latitude_stats}
\end{figure}

We quantify the boulder abundance in three latitude zones (low, mid, and high) on both Ceres and Vesta in Fig.~\ref{fig:latitude_stats}. We calculated both the density of boulders and the density of craters with at least one boulder larger than 105~m by dividing the total number of boulders/craters in a latitudinal zone by the total surface area of the zone (including shadowed terrain), calculated under the assumption that Ceres and Vesta are spheres with radii of 469 and 263~km, respectively. Adopting Poisson error bars allows to assess whether any differences are the result of chance. The boulder density graph (Fig.~\ref{fig:latitude_stats}a) confirms our visual impression that the density is higher at the high latitudes of Ceres, despite the fact that polar terrain was partly in shadow. The small (Poisson) error bars indicate that this is very likely not due to chance. The Ceres boulder density at mid-latitudes is also a little higher than that at low latitudes. The boulder density at low latitudes is more uncertain than the error bar indicates. First, boulder numbers may be underestimated because of the limited visibility of the shadows cast by smaller boulders (see Sec.~\ref{sec:mapping}). Second, boulder counts for Occator and Haulani, with their large scale flows, are uncertain (see Sec.~\ref{sec:general_stats}). In contrast to Ceres, the Vesta boulder density does not vary with latitude, within the Poisson error margins. It is a little lower than the Ceres boulder density at low and mid-latitudes and much lower at high latitudes. The density of craters-with-boulders (Fig.~\ref{fig:latitude_stats}b) also increases on Ceres toward high latitudes, although the correlation is not as strong due to the larger error bars. The density of craters-with-boulders on Vesta does not vary with latitude, within the error margins. The crater density may be similar at high latitudes on both bodies, although the error bars are large. Interestingly, the density of craters-with-boulders on Vesta is significantly higher than that on Ceres for low and mid-latitudes, despite the boulder density being lower. This implies that, on average, craters on Ceres have more boulders than on Vesta, consistent with Fig.~\ref{fig:general_stats}a.

\begin{figure}
	\centering
	\includegraphics[width=\textwidth,angle=0]{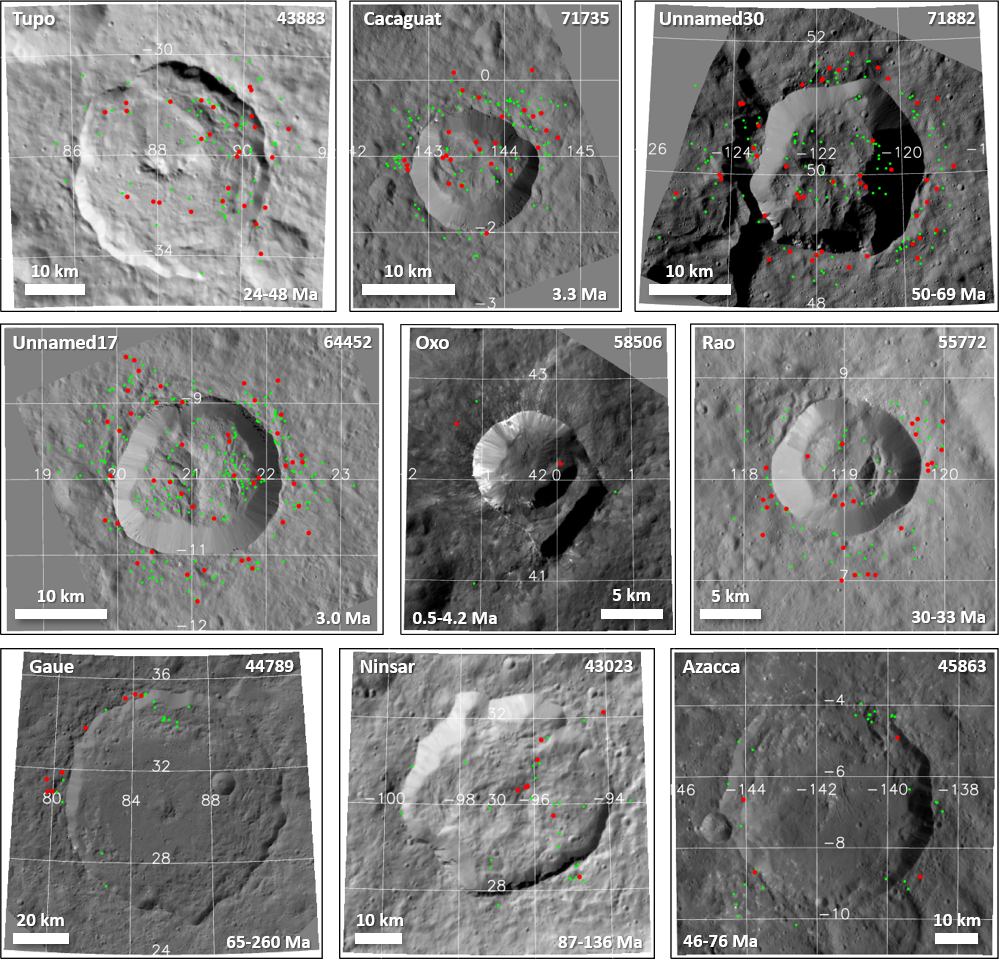}
	\caption{Spatial boulder distribution for several craters for which an age estimate is available. Green, small dots represent boulders with a size between 3 and 4~pixels ($105$~m~$< d < 140$~m). Red, large dots represent boulders larger than 4~pixels ($d > 140$~m). The FC2 image number is indicated in the top right.}
	\label{fig:craters}
\end{figure}

\begin{figure}
	\centering
	\includegraphics[width=\textwidth,angle=0]{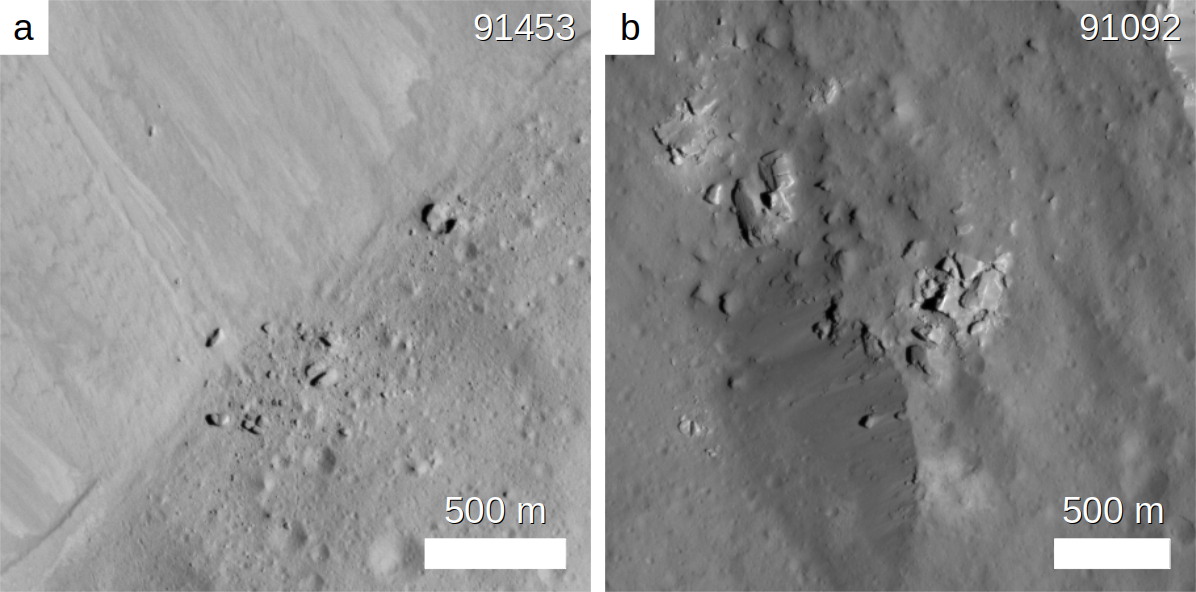}
	\caption{High-resolution (full-size) images of boulders from the extended mission phase. {\bf a}.~Boulders collected at the foot of the wall of Occator crater. Image center is at (lat, lon) $= (13.8^\circ, 241.3^\circ)$. {\bf b}.~Boulders that appear to have fragmented in place, either through impact or erosion. Image center is at (lat, lon) $= (11.2^\circ, 244.0^\circ)$. The FC2 image number is indicated in the top right.}
	\label{fig:XM2}
\end{figure}

Examples\footnote{Maps of the boulder distribution around all craters in Table~\ref{tab:craters} are available for download (see data availability statement).} of the spatial distribution of boulders around individual craters are shown in Fig.~\ref{fig:craters}. Boulders are located both inside the crater and outside the rim, typically, all within one crater radius. The figure distinguishes between boulders in two different size classes, but sorting of boulders according to their size is not evident, consistent with the findings of \citet{S18}. There are no accumulations of boulders with a size range considered in our global study at the foot of steep slopes, which argues against a formation of such boulders by post-impact weathering. High-resolution images (scales of 3-10~m/px) acquired in the Dawn extended mission show evidence for boulder transport on crater walls, such as bounce marks on unconsolidated talus and boulders at the downslope end of tracks. Figure~\ref{fig:XM2}a shows an example of boulders collected at the foot of a crater wall. Such boulders are consistently smaller than those we consider here. As very high-resolution images are only available for a very small fraction of the surface, we do not consider them for our global study. We therefore decided to group boulders inside and outside the craters together and treat them as a single population. Another image from the extended mission, in Fig.~\ref{fig:XM2}b, shows clusters of boulders that appear to derive from former, larger boulders. Such fields of debris may originate from either impact of the larger boulder on the surface or weathering and/or erosion in place, demonstrating that boulders disintegrate by fracturing.

\subsection{Boulder lifetime}
\label{sec:age}

Boulders degrade over time and eventually disappear from the surface. \citet{B15} predicted that the survival time of meter-sized boulders on Ceres is very similar to that on Vesta, based on estimates of the potential impactor flux and the expected impact velocities. \citet{S20} determined a survival time of about 300~Ma for Vesta boulders, much larger than the $\sim 10$~Ma predicted by \citet{B15}. The authors attributed this apparent discrepancy to the fact that the boulders in their sample were one to two orders of magnitude larger than the meter-scale associated with the prediction. The boulders in our sample are even larger than the Vesta boulders studied by \citet{S20} because of the lower {\sc lamo} image resolution at Ceres, but typically by a factor of two rather than an order of magnitude. It should therefore be possible to test the \citet{B15} prediction of similar survival times on Vesta and Ceres, if accurate age estimates are available for our Ceres craters.

Estimating the age of a crater is typically done by counting smaller craters on a selected area on the ejecta blanket and modeling the resulting SFD. Just as for Vesta, two alternative chronologies have been used to model crater SFDs for Ceres: The lunar-derived model (LDM) adapts the lunar production and chronology functions to impact conditions on Ceres, whereas the asteroid-derived model (ADM) derives a production function by scaling the observed SFD from the main asteroid belt to the SFD of Ceres craters \citep{H16}. Most papers on the topic of Ceres dating employ both chronologies and provide two age estimates for a particular crater. Table~\ref{tab:ages} lists craters for which age estimates are available. The uncertainty associated with the age is typically large, as the two chronologies can yield widely different values. Additional sources of uncertainty are the choice of counting area and the assumed strength of the target surface. The tabulated ages were estimated assuming an impact into hard rock. \citet{W18} found that the ADM (and, presumably, the LDM) ages are much larger for a rubble surface (e.g.\ Cacaguat: 14.4 instead of 3.3~Ma, Rao: 133 instead of 30.4~Ma).

\begin{figure}
	\centering
	\includegraphics[width=\textwidth,angle=0]{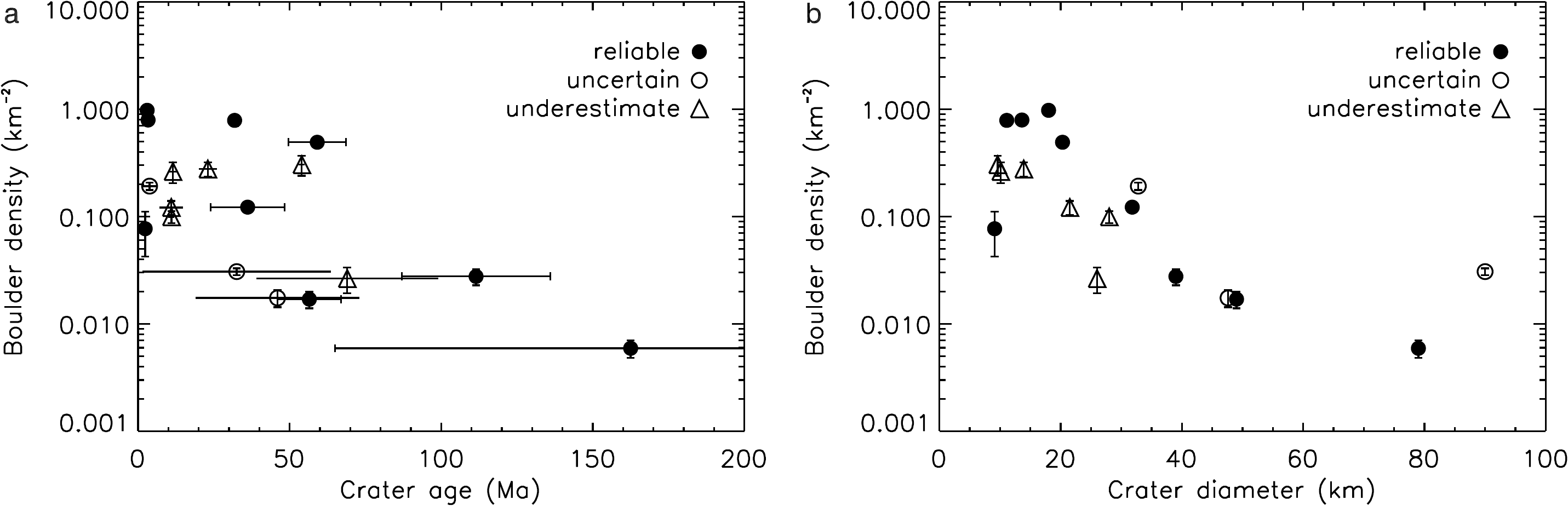}
	\caption{The areal density of boulders larger than 3~pixels ($d > 105$~m) (Table~\ref{tab:ages}). {\bf a}.~Density versus crater age. {\bf b}.~Density versus crater diameter. The error bars on the density were calculated assuming the number of boulders follows a Poisson distribution. The open symbols represent craters whose boulder density is unreliable, either because of uncertain boulder identifications or because the associated craters were partly in the shadow (underestimate).}
	\label{fig:boulder_density}
\end{figure}

We define ``areal density'' as the total number of boulders identified in and around a crater divided by the crater equivalent area, calculated as the area of a circle with the diameter for that crater (Table~\ref{tab:craters}). We determined the areal density of boulders larger than 105~m in and around the craters in Table~\ref{tab:ages}. Some areal densities are unreliable. Boulder numbers are underestimated for craters at high latitudes, which were partly in the shadow during {\sc lamo} (Shennong, Unnamed4/26/28/36/44). Boulder numbers are uncertain for craters that show post-impact modifications in the form of flows (Haulani, Ikapati, Occator). The distribution of boulders around craters with a reliable density was shown in Fig.~\ref{fig:craters}. We relate the boulder density to crater age in Fig.~\ref{fig:boulder_density}a, where the age ranges are spanned by the LDM and ADM estimates. The two variables are anti-correlated. There is a large degree of scatter in the data, which may be due to the fact that the boulder density also depends on latitude (see Sec.~\ref{sec:global_distribution}). The data suggest that the maximum boulder survival time is around 150~Ma, where we note that the age of the oldest crater in the figure (Gaue) is very uncertain. Support for this maximum age comes from craters of this age for which we did not find boulders: One such, unnamed, crater at ($162^\circ$E, $+78^\circ$) has an estimated age of 89-252~Ma \citep{R18}. Two others, Messor at ($234^\circ$E, $+50^\circ$) and an unnamed crater at ($186^\circ$E, $+23^\circ$), have estimated ages of 96-192~Ma and 88-205~Ma, respectively \citep{SB18}. All craters estimated to be younger than 150~Ma in the papers referenced in Table~\ref{tab:ages} have boulders. Given the uncertainty in the crater age estimates due to the different models, we adopt an uncertainty of 50~Ma for the boulder survival time. Whereas the correlation with crater age is expected, we also find that the boulder density is correlated with crater size (Fig.~\ref{fig:boulder_density}b). This may at least partly be explained by the fact that large craters are, on average, older than small craters. Another explanation might be that larger craters sample different, deeper crustal layers in a stratified crust, a concept that we discuss in the next section.

\citet{B15} estimated the survival time of meter-sized boulders on Vesta, Ceres, and the Moon by predicting the impactor velocity and density distributions. Boulders on Vesta and Ceres should live equally long, but 30 times shorter than on the Moon. \citet{S20} found that the maximum boulder survival time on Vesta is 300~Ma, the same as that of lunar boulders \citep{B13}. The authors attributed this apparent contradiction to the fact that the Vesta boulders in their study are larger than 60~m, rather than meter-sized. In other words, large boulders live longer than small ones. At $150 \pm 50$~Ma, the survival time of the Ceres boulders in our sample is only half that of Vesta boulders, despite being larger ($> 105$~m). Thus, the lifetime of Ceres boulders may be less than half that of Vesta boulders of the same size.

\section{Discussion}
\label{sec:discussion}

In the previous section we described the properties of the population of large boulders on Ceres and compared them to those of the Vesta population. Our major findings are: (1)~Ceres craters have, on average, more boulders than Vesta craters of the same size, (2)~the largest boulders are, on average, somewhat larger for Ceres craters than Vesta craters of the same size, (3)~the SFD of the global boulder population is better described by a Weibull distribution than a power law for both Ceres and Vesta, (4)~boulders on Ceres are more numerous at high latitudes than at mid- and low latitudes, in contrast to Vesta, and (5)~boulders live shorter on Ceres than on Vesta. How can we reconcile these findings? Let us start with finding~(3), which supports the idea that the SFD of particles on the surface of small bodies follows a Weibull distribution rather than a power law \citep{S20}. Unfortunately, uncertainties regarding the boulder populations of the largest craters on both Ceres and Vesta prevent us from a meaningful comparison of the Weibull parameters.

To address the other findings, we need to consider how boulders degrade. The dominant mechanism responsible for degradation are weathering due to thermal stress \citep{D14,M17,EM19} and meteorite impacts \citep{B20}. The efficacy of weathering due to diurnal thermal cycling or thermal shock correlates with the rate of surface temperature change ($dT/dt$). Rapid temperature changes occur at sunrise and sunset and during daytime shadowing. On quickly rotating bodies such as Vesta and Ceres, sunrise and sunset are the main drivers of $dT/dt$ \citep{MB12}. Vesta and Ceres have rotation rates of 5.34~h and 9.07~h, so $dT/dt$ at the terminator should be larger on Vesta. Moreover, the thermal cycling rate is higher on Vesta, and a boulder of the same age will have experienced more thermal cycles than on Ceres. Therefore, boulders of identical lithology would degrade faster on Vesta through thermal stress weathering. This is inconsistent with the shorter boulder lifetime on Ceres (finding~5), but consistent with the fact that boulders for a given crater diameter are larger on Ceres (finding~2). It is a different story for boulder degradation due to meteorite impacts. Because of their similar locations in the main asteroid belt, Ceres and Vesta experience similar impact regimes, in terms of impactor size distribution, flux, and velocity \citep{B15}. Therefore, boulders of identical lithology would degrade equally fast on both worlds through meteorite impacts.

The crusts of Vesta and Ceres are of different composition, leading to differences in the compressive and tensile strengths that control the resistance to stress. Vesta's crust is mostly an assemblage of eucritic basalts and pyroxene cumulates \citep{dS12}, whereas Ceres has a likely carbonaceous chondrite bulk composition \citep{P17,McS18}, with widespread phyllosilicates \citep{A16}, localized deposits of salts \citep{dS16}, and water ice present just meters below the surface \citep{P17,SH17}. On average, the materials in Ceres' crust are mechanically weaker than those in Vesta's crust, so its boulders are less resistant to stress. This applies to both thermal stress and stress caused by meteorite impacts. Given the considerable uncertainties in quantifying the effects of thermal stresses and determining thermal strain thresholds (e.g., \citealt{BJ13}), it is impossible to predict which of these competing effects dominates (higher thermal stress experienced on Vesta or lower resistance to stress on Ceres). Nevertheless, our finding~(5) that boulders on Ceres have a shorter lifetime suggests that the lower mechanical strength of the Ceres crust is primarily responsible. The lower crustal strength would also lead to the formation of a larger crater on Ceres than Vesta for an identical impactor. So a crater of the same diameter is, on average, younger on Ceres than on Vesta. Boulders disappear over time, and younger craters have, on average, more boulders than older craters of the same size. Therefore, craters of the same size are expected to have more boulders on Ceres than Vesta, explaining finding~(1).

The prevalence of large boulders at high latitudes may be explained by a higher rate of physical weathering and boulder breakdown at lower latitudes as compared to higher latitudes. Ceres has a obliquity of only $4^\circ$ at present\footnote{The obliquity of Ceres varied between $2^\circ$ and $20^\circ$ over the last 20~Ma \citep{E17,V19}, so the latitudinal difference in thermal stress may not always have been as high as today. Given the lifetime of boulders as derived in this study (150~Ma), the present situation is not necessarily representative.} \citep{E17}, and therefore the diurnal temperature waves are expected to be larger at lower latitudes \citep{HA15}. The duration of sunrise and sunset would also be shorter at lower latitudes, increasing $dT/dt$. Both effects would lead to relatively faster boulder breakdown by thermal stresses at equatorial latitudes, consistent with finding~(4). As water ice is likely abundant in the subsurface, Ceres boulders may harbor a significant fraction of ice. Ice would be relatively stable inside these large boulders, just as it is stable just meters below the surface \citep{FS89}, yet ice-rich boulders could be more prone to degradation by thermal stress, as fractures may be widened by sublimation, further weakening the boulder structure. The hypothesis that Ceres' boulders are rich in water ice is consistent with \citet{Ri14}, who considered the question why Ceres does not have a dynamical family. Large impacts on Ceres would produce escaping fragment asteroids (essentially liberated boulders), but the absence of a dynamical family led the authors to suggest that such escaped fragments are ice-rich and prematurely destroyed by sublimation.

To identify other factors that may contribute to latitudinal differences in the areal densities of boulders and craters with boulders, we consider the independent global data set of floor-fractured craters. Craters with fractured floors are indicators for several possible processes, including updoming due to cryomagmatic activity, as for instance beneath Occator crater \citep{B18}. Cryomagmatism may indicate the presence of a crustal column beneath floor-fractured craters that is enriched in volatiles and therefore mechanically weaker. A volatile-rich and weak target material is expected to eject boulders with these properties, which would be more susceptible to degradation. Twenty-one floor-fractured craters have been identified on Ceres \citep{BS18}, seven of which are marked in Table~\ref{tab:ages}. Floor-fractured craters tend to have low boulder densities, consistent with the expectation that boulders ejected from floor-fractured craters degrade faster, resulting, on average, in a lower boulder density. However, the floor-fracturing is likely related to post-impact processes, and it is not clear whether the target substrate was already weaker before the impact.

Support for the hypothesis of a crust with a mechanical strength that depends on latitude comes from the observation that five of the seven floor-fractured craters in Table~\ref{tab:ages} (Azacca, Haulani, Ikapati, Occator, and Tupo) display concentric fracturing beyond the crater rim, suggestive of creep of a low-viscosity, and therefore mechanically weak, subsurface layer \citep{O19}. Such craters with concentric fractures are mostly located between latitudes $46^\circ$S and $34^\circ$N. This is consistent with landslide morphology that suggests the presence of a relatively weak layer at low- to mid-latitudes \citep{C19}. This layer thins towards the poles and overlies a stronger layer, in agreement with lower temperatures at high latitudes increasing crustal viscosity and strength \citep{B16}. Boulders excavated from the weaker layer would tend to degrade faster, and more boulders would be retained in the polar regions. Moreover, the subsurface ice content is lower in low- to mid-latitudes regions \citep{SH17}, hence boulders there would harbor more impurities like phyllosilicates, which tend to lower the albedo, raise temperatures, and enhance sublimation \citep{Ri14}. This effect would lead to faster degradation of low- and mid-latitude, less ice-rich boulders as compared with polar boulders with possibly a higher ice content. Therefore, there may be a causal relationship between the boulder density and the pre-impact properties of the crust. So, while boulders may harbor water ice, the complexity of Ceres' crust, with its laterally and vertically varying properties \citep{B16,O19,P20}, precludes any definite conclusion on the observed distribution of boulder and boulder crater densities across the surface.

\section*{Data availability}

Dawn framing camera images are available from NASA's Planetary Data System at \url{https://pds.nasa.gov/}. Our Ceres boulder data, including maps of the boulder distribution around all craters, are available for download at \url{https://doi.org/10.5281/zenodo.4715154}.

\section*{Acknowledgments}

We are grateful for technical support provided by J-Ceres developer Dale Noss and his team at ASU. We thank Maurizio Pajola and an anonymous reviewer for helpful suggestions to improve the manuscript.


\bibliography{Boulders}
\bibliographystyle{aasjournal}


\newpage
\clearpage

\begin{table}
	\centering
	\caption{All craters on Ceres with at least one boulder larger than 4~pixels ($d > 140$~m). Crater and boulder diameters are $D$ and $d$, respectively, and $\alpha$ is the power law exponent of the (cumulative) boulder SFD as derived with the ML method (only for craters with $n_{d > 4 {\rm px}} > 5$).}
	\label{tab:craters}
	\begin{tabular}{lllllllll}
		\hline
		Name & Longitude & Latitude & $D$ & $d_{\rm max}$ & $n_{d > 3 {\rm px}}$ & $n_{d > 4 {\rm px}}$ & $\alpha$ & $\sigma_\alpha$ \\
		& ($^\circ$E) & ($^\circ$) & (km) & (m) & & & & \\
		\hline
		Azacca & 218.4 & $-6.6$ & 49.0 & 171 & 32 & 4 & & \\
		Braciaca & 84.3 & $-22.7$ & 7.7 & 179 & 17 & 1 & & \\
		Cacaguat & 143.6 & $-1.2$ & 13.6 & 353 & 115 & 26 & $-4.9$ & 1.0 \\
		Cozobi & 287.3 & $+45.3$ & 22.4 & 182 & 28 & 6 & $-8.5$ & 3.5 \\
		Emesh & 158.2 & $+11.1$ & 18.0 & 214 & 43 & 9 & $-7.1$ & 2.4 \\
		Gaue & 86.2 & $+30.8$ & 79.0 & 254 & 29 & 8 & $-3.9$ & 1.4 \\
		Haulani & 10.9 & $+5.7$ & 32.8 & 279 & 164 & 20 & $-7.5$ & $1.7$ \\
		Ialonus & 168.5 & $+48.2$ & 15.9 & 205 & 29 & 10 & $-7.7$ & 2.4 \\
		Ikapati & 45.6 & $+33.8$ & 47.6 & 155 & 31 & 5 & & \\
		Jacheongbi & 2.3 & $-69.2$ & 26.8 & 498 & 473 & 160 & $-4.5$ & 0.4 \\
		Juling & 168.5 & $-35.9$ & 17.3 & 202 & 85 & 22 & $-6.8$ & 1.5 \\
		Kokopelli & 124.5 & $+18.3$ & 31.2 & 256 & 264 & 93 & $-4.6$ & 0.5 \\
		Kupalo & 173.5 & $-39.4$ & 25.1 & 231 & 81 & 21 & $-5.7$ & 1.2 \\
		Ninsar & 263.3 & $+30.3$ & 39.0 & 190 & 33 & 8 & $-7.7$ & 2.7 \\
		Nunghui & 272.3 & $-54.0$ & 21.9 & 287 & 284 & 84 & $-5.1$ & 0.6 \\
		Occator & 239.4 & $+19.7$ & 90.0 & 229 & 195 & 26 & $-7.0$ & 1.4 \\
		Oxo & 359.6 & $+42.2$ & 9.1 & 172 & 5 & 2 & & \\
		Rao & 119.0 & $+8.1$ & 11.1 & 338 & 76 & 25 & $-5.2$ & 1.0 \\
		Ratumaibulu & 77.7 & $-67.3$ & 18.2 & 219 & 98 & 23 & $-6.7$ & 1.4 \\
		Sekhet & 255.8 & $-66.4$ & 38.0 & 259 & 167 & 59 & $-5.9$ & 0.8 \\
		Shennong & 28.1 & $+69.0$ & 28.0 & 187 & 61 & 14 & $-6.5$ & 1.7 \\
		Tawals & 238.0 & $-39.1$ & 7.9 & 160 & 15 & 2 & & \\
		Thrud & 31.1 & $-71.3$ & 6.5 & 140 & 18 & 1 & & \\
		Tupo & 88.4 & $-32.4$ & 31.8 & 218 & 97 & 27 & $-6.5$ & 1.3 \\
		Unnamed02 & 10.8 & $-2.7$ & 6.6 & 219 & 52 & 14 & $-6.3$ & 1.7 \\
		Unnamed03 & 37.1 & $+34.8$ & 6.5 & 221 & 10 & 2 & & \\
		Unnamed04 & 58.2 & $+58.6$ & 10.1 & 214 & 21 & 2 & & \\
		Unnamed06 & 86.4 & $-10.7$ & 5.3 & 225 & 18 & 3 & & \\
		Unnamed09 & 270.1 & $+50.2$ & 8.7 & 174 & 16 & 3 & & \\
		Unnamed11 & 279.1 & $-23.0$ & 15.0 & 386 & 209 & 74 & $-4.7$ & 0.5 \\
		Unnamed12 & 307.0 & $-35.7$ & 7.0 & 188 & 15 & 7 & $-7.5$ & 2.9 \\
		Unnamed14 & 344.1 & $+18.1$ & 9.3 & 248 & 35 & 11 & $-6.9$ & 2.1 \\
		Unnamed15 & 244.4 & $-67.2$ & 9.3 & 180 & 33 & 7 & $-8.0$ & 3.0 \\
		Unnamed16 & 266.4 & $-34.3$ & 13.3 & 267 & 181 & 46 & $-6.8$ & 1.0 \\
		Unnamed17 & 21.1 & $-10.0$ & 18.0 & 213 & 248 & 45 & $-7.9$ & 1.2 \\
		Unnamed18 & 355.7 & $+33.6$ & 8.0 & 174 & 26 & 6 & $-7.7$ & 3.1 \\
		Unnamed19 & 153.1 & $-68.5$ & 11.7 & 150 & 30 & 7 & $-18.2$ & 6.9 \\
		Unnamed21 & 348.6 & $-3.5$ & 7.4 & 179 & 24 & 8 & $-8.3$ & 3.0 \\
		Unnamed22 & 62.4 & $+17.0$ & 8.8 & 167 & 22 & 4 & & \\
		Unnamed23 & 232.0 & $-64.7$ & 6.0 & 156 & 11 & 1 & & \\
		\hline
	\end{tabular}
\end{table}

\begin{table}
	\centering
	\begin{tabular}{lllllllll}
		\hline
		Name & Longitude & Latitude & $D$ & $d_{\rm max}$ & $n_{d > 3 {\rm px}}$ & $n_{d > 4 {\rm px}}$ & $\alpha$ & $\sigma_\alpha$ \\
		& ($^\circ$E) & ($^\circ$) & (km) & (m) & & & & \\
		\hline
		Unnamed24 & 312.7 & $-45.0$ & 11.5 & 247 & 33 & 7 & $-5.0$ & 1.9 \\
		Unnamed26 & 252.6 & $+62.9$ & 21.5 & 237 & 44 & 12 & $-8.1$ & 2.3 \\
		Unnamed27 & 191.6 & $-25.8$ & 10.9 & 191 & 46 & 7 & $-7.5$ & 2.8 \\
		Unnamed28 & 46.7 & $+79.7$ & 13.9 & 176 & 42 & 10 & $-9.9$ & 3.1 \\
		Unnamed29 & 199.1 & $-11.4$ & 15.3 & 179 & 57 & 15 & $-8.7$ & 2.2 \\
		Unnamed30 & 238.5 & $+50.0$ & 20.3 & 237 & 159 & 45 & $-8.2$ & 1.2 \\
		Unnamed31 & 276.5 & $-5.0$ & 5.0 & 153 & 20 & 1 & & \\
		Unnamed32 & 300.8 & $+68.6$ & 6.8 & 150 & 20 & 4 & & \\
		Unnamed33 & 19.6 & $-47.9$ & 7.6 & 179 & 27 & 4 & & \\
		Unnamed34 & 111.0 & $-39.4$ & 18.6 & 248 & 112 & 33 & $-6.8$ & 1.2 \\
		Unnamed36 & 86.5 & $+58.8$ & 9.6 & 215 & 22 & 4 & & \\
		Unnamed37 & 332.3 & $-52.9$ & 10.3 & 265 & 35 & 7 & $-4.5$ & 1.7 \\
		Unnamed42 & 131.7 & $+65.7$ & 19.4 & 188 & 35 & 8 & $-6.0$ & 2.1 \\
		Unnamed43 & 134.1 & $+68.0$ & 15.1 & 266 & 68 & 18 & $-5.8$ & 1.4 \\
		Unnamed44 & 350.0 & $+68.2$ & 26.0 & 226 & 14 & 5 &  &  \\
		Unnamed48 & 162.1 & $-18.0$ & 8.9 & 143 & 2 & 1 &  &  \\
		Victa & 301.0 & $+36.2$ & 28.9 & 184 & 39 & 12 & $-8.9$ & 2.6 \\
		Xevioso & 310.6 & $+0.7$ & 8.3 & 216 & 16 & 3 & & \\
		\hline
	\end{tabular}
\end{table}

\begin{table}
	\centering
	\caption{Age and areal boulder density for craters for which an age estimate is available. Age ranges are spanned by the LDM and ADM estimates, ignoring the associated standard deviations. Density is defined as the total number of boulders larger than 3~pixels divided by crater equivalent area. Bracketed densities are uncertain.}
	\begin{tabular}{lllll}
		\hline
		Name & Age & Area& Density & Source for age \\
		& (Ma) & ($10^3$ km$^2$) & (km$^{-2}$) & \\
		\hline
		Azacca\tablenotemark{a} & 46-76 & 1.89 & $0.017 \pm 0.003$ & \citet{S16} \\
		Cacaguat & 3.3 & 0.145 & $0.79 \pm 0.07$ & \citet{W18} \\
		Gaue\tablenotemark{a} & 65-162 & 4.90 & $0.0059 \pm 0.0011$ & \citet{S16} \\
		& 69-260 & & & \citet{P18} \\
		Haulani\tablenotemark{a} & 1.7-5.9 & 0.845 & ($0.19 \pm 0.02$) & \citet{S16} \\
		& 1.7-2.6 & & & \citet{K18} \\
		Ikapati\tablenotemark{a} & 19-66 & 1.78 & ($0.017 \pm 0.003$) & \citet{S16} \\
		& 36-73 & & & \citet{P18} \\
		Ninsar\tablenotemark{a} & 87-136 & 1.19 & $0.028 \pm 0.005$ & \citet{SB18} \\
		Occator\tablenotemark{a} & 1.4-64 & 6.36 & ($0.031 \pm 0.002$) & \citet{N19} \\
		Oxo & 0.5 & 0.0650 & $0.08 \pm 0.03$ & \citet{S16} \\
		& 3.7-4.2 & & & \citet{H18} \\
		Rao & 30-33 & 0.0968 & $0.79 \pm 0.09$ & \citet{W18} \\
		Shennong & 10-12 & 0.616 & $> 0.10 \pm 0.01$ & \citet{R18} \\
		Tupo\tablenotemark{a} & 36-48 & 0.794 & $0.12 \pm 0.01$ & \citet{S16} \\
		& 24-29 & & & \citet{SK18} \\
		Unnamed4 & 10-13 & 0.0801 & $> 0.26 \pm 0.06$ & \citet{P18} \\
		Unnamed17 & 3.0 & 0.254 & $0.97 \pm 0.06$ & \citet{S16} \\
		Unnamed26 & 7.0-15 & 0.363 & $> 0.12 \pm 0.02$ & \citet{SB18} \\
		Unnamed28 & 20-26 & 0.152 & $> 0.28 \pm 0.04$ & \citet{R18} \\
		Unnamed30 & 50-69 & 0.324 & $0.49 \pm 0.04$ & \citet{SB18} \\
		Unnamed36 & 53-55 & 0.0724 & $> 0.30 \pm 0.06$ & \citet{P18} \\
		Unnamed44 & 39-99 & 0.531 & $> 0.026 \pm 0.007$ & \citet{R18} \\
		\hline
	\end{tabular}
\tablenotetext{a}{Floor-fractured crater.}
	\label{tab:ages}
\end{table}

\newpage
\clearpage

\end{document}